\documentclass[11pt]{article}
\usepackage{putex}
\usepackage{graphicx}
\usepackage{latexsym,amsmath,amsfonts,amssymb}
\usepackage{bbm}
\usepackage{mathtools}
\usepackage{hyperref}

\renewenvironment{thebibliography}[1]{%
\begin{oldthebibliography}{#1}%
\setlength\itemsep{-2mm}%
}%
{%
\end{oldthebibliography}%
}

\newcommand{\trans}{\ensuremath{\mathsf T}}
\newcommand{\dvol}{d\mathrm{vol}}

\newcommand{\du}[2]{_{#1}^{\phantom{#1}#2}}
\newcommand{\rep}[1]{\ensuremath{\mathbf{#1}}}

\newcommand{\eg}{\textit{e.g.}}

\newcommand{\ie}{\textit{i.e.}}

\numberwithin{equation}{section}

\newcommand{\nn}{\nonumber}
\newcommand{\mat}[1]{\begin{pmatrix} #1 \end{pmatrix}}
\newcommand{\smat}[1]{\big( \begin{smallmatrix} #1 \end{smallmatrix} \big)}
\newcommand{\be}{\begin{equation}} \newcommand{\ee}{\end{equation}}
\newcommand{\bea}{\begin{equation} \begin{aligned}} \newcommand{\eea}{\end{aligned} \end{equation}}

\newcommand{\cL}{\mathcal{L}}

\newcommand{\cN}{\mathcal{N}}
\newcommand{\cO}{\mathcal{O}}

\newcommand{\cQ}{\mathcal{Q}}
\newcommand{\cR}{\mathcal{R}}

\newcommand{\cW}{\mathcal{W}}

\newcommand{\bR}{\mathbb{R}}
\newcommand{\bZ}{\mathbb{Z}}

\newcommand{\fg}{\mathfrak{g}}
\newcommand{\fm}{\mathfrak{m}}
\newcommand{\fn}{\mathfrak{n}}
\newcommand{\fq}{\mathfrak{q}}
\newcommand{\fR}{\mathfrak{R}}
\newcommand{\fS}{\mathfrak{S}}

\newcommand{\fz}{\mathfrak{z}}
\newcommand{\unit}{\mathbbm{1}}

\DeclareMathOperator{\Tr}{Tr}

\DeclareMathOperator{\rank}{rank}
\DeclareMathOperator*{\Res}{Res}
\DeclareMathOperator{\re}{\mathbb{R}e}

\DeclareMathOperator{\ind}{ind}

\begin{document}


\title{Higgs branch localization in three dimensions}

\authors{Francesco Benini$^\diamondsuit$ and Wolfger Peelaers$^\spadesuit$}

\institution{SB}{${}^\diamondsuit$
Simons Center for Geometry and Physics, Stony Brook University, \cr
$\;\:\,$ Stony Brook, NY 11794, USA}
\institution{YITP}{${}^\spadesuit$
C.N. Yang Institute for Theoretical Physics, Stony Brook University,  \cr
$\;\;\,$ Stony Brook, NY 11794, USA}

\abstract{We show that the supersymmetric partition function of three-dimensional $\cN=2$ R-symmetric Chern-Simons-matter theories on the squashed $S^3$ and on $S^2 \times S^1$ can be computed with the so-called Higgs branch localization method, alternative to the more standard Coulomb branch localization. For theories that could be completely Higgsed by Fayet-Iliopoulos terms, the path integral is dominated by BPS vortex strings sitting at two circles in the geometry. In this way, the partition function directly takes the form of a sum, over a finite number of points on the classical Coulomb branch, of a vortex-string times an antivortex-string partition functions.}

\preprint{YITP-SB-13-46}

\maketitle


{
\setcounter{tocdepth}{2}
\setlength\parskip{-0.7mm}
\tableofcontents
}

\section{Introduction}
\label{sec: intro}

In the last few years there has been a huge development in the study of supersymmetric quantum field theories on compact manifolds, without topological twist. A stunning feature is that, in many cases, we are able to compute \emph{exactly} the path integral and the expectation values of (local and non-local) operators that preserve some supersymmetry, with localization techniques \cite{Witten:1988ze, Witten:1991zz}. The path integral can be reduced to something much simpler, like a matrix integral or a counting problem, and explicitly evaluated. After the seminal work of Pestun on $S^4$ \cite{Pestun:2007rz}, the techniques have been developed in many different contexts, essentially from two to five dimensions (see \cite{Benini:2012ui, Doroud:2012xw, Gadde:2013dda, Benini:2013nda, Benini:2013xpa, Kapustin:2009kz, Jafferis:2010un, Hama:2010av, Hama:2011ea, Imamura:2011wg, Alday:2013lba, Kim:2009wb, Imamura:2011su, Pestun:2007rz, Hama:2012bg, Closset:2013sxa, Kallen:2012cs, Hosomichi:2012ek, Kim:2012ava, Imamura:2012xg} for a non-exhaustive list).

Most of the work on supersymmetric theories with no twisting has been within the so-called \emph{Coulomb branch localization}: the path integral is reduced to an ordinary integral over a ``classical Coulomb branch'',%
\footnote{We used quotation marks because that would be the classical Coulomb branch on flat space, while the theories we consider are on compact Euclidean curved manifolds.}
parametrized either by scalars in the vector multiplets, or by holonomies around circles. The integrand can contain non-perturbative contributions (\eg{} if the geometry contains an $S^4$ or $S^5$), or not. For instance, in three dimensions \cite{Kapustin:2009kz, Jafferis:2010un, Hama:2010av, Hama:2011ea, Imamura:2011wg, Kim:2009wb, Imamura:2011su} the integrand is simply the one-loop determinant of all fields around the Coulomb branch configurations. It was observed by S.~Pasquetti \cite{Pasquetti:2011fj} (inspired by \cite{Dimofte:2011ju}), though, that the $S^3$ partition function can be rewritten as a sum over a finite set of points on the Coulomb branch, of the vortex times the antivortex partition functions \cite{Shadchin:2006yz},%
\footnote{The vortex partition function counts vortices in the $\Omega$-background on $\bR^2$, in the same way as the instanton partition function of \cite{Nekrasov:2002qd} counts instantons on $\bR^4$.}
which do have a non-perturbative origin. In this paper we would like to gain a better understanding of this phenomenon, from the point of view of localization.

A mechanism responsible for such a ``factorization'' was first understood in \cite{Benini:2012ui, Doroud:2012xw}, in the analogous context of 2d $\cN=(2,2)$ theories on $S^2$. It is possible to perform localization in an alternative way (that can be thought of either as adding a different deformation term, or as choosing a different path integration contour in complexified field space), dubbed \emph{Higgs branch localization}, such that the BPS configurations contributing to the path integral are vortices at the north pole and antivortices at the south pole of $S^2$. Notice that such 2d factorization for supersymmetric non-twisted theories is tightly related to the more general $tt^*$ setup \cite{Cecotti:1991me}.

In three dimensions quite some work has been done to understand factorization. Building on \cite{Dimofte:2011ju}, the authors of \cite{Beem:2012mb} gave very general arguments why factorization should take place in terms of ``holomorphic blocks''. Factorization has been explicitly checked for $U(N)$ theories with (anti)fundamentals on $S^3$ \cite{Taki:2013opa} and $S^2 \times S^1$ \cite{Krattenthaler:2011da, Hwang:2012jh},%
\footnote{We slightly revisit some manipulations in \cite{Taki:2013opa, Hwang:2012jh}.}
manipulating the Coulomb branch integrals. General continuous deformations of the geometry have been studied in \cite{Alday:2013lba, Closset:2013vra}. Finally, the more general $tt^*$ setup has been developed in three and four dimensions \cite{Cecotti:2013mba}. Our approach is different.

In this paper we are after a \emph{Higgs branch localization} mechanism in three-dimensional $\cN=2$ R-symmetric Chern-Simons-matter theories, similar to the two-dimensional one \cite{Benini:2012ui}. We focus on the squashed sphere $S^3_b$ and on $S^2\times S^1$, knowing that more general backgrounds could be analyzed with the tools of \cite{Klare:2012gn, Closset:2012ru, Closset:2013vra}. We show that both on $S^3_b$ and $S^2\times S^1$, as in \cite{Benini:2012ui}, an alternative localization (based on a different deformation term) is possible which directly yields an expression
$$
Z = \sum_\text{vacua} Z_\text{cl} \; Z'_\text{1-loop} \; Z_\text{v} \; Z_\text{av} \;,
$$
whenever the flat-space theory could be completely Higgsed by a Fayet-Iliopoulos term, and with some bounds on the Chern-Simons levels which apparently have been overlooked before. The sum is over a finite set of points on the would-be ``Coulomb branch'', where some chiral multiplets get a VEV solving the D-term equations and completely Higgsing the gauge group. What is summed is a classical and one-loop contribution, evaluated on the vacua, times a vortex and an antivortex contributions, coming from BPS vortex-strings at the northern and southern circles of $S^3_b$ or $S^2\times S^1$. Both can be expressed in terms of the \emph{vortex partition function} (VPF) on the twisted $\bR^2_\epsilon \times S^1$ (a version of the VPF on the $\Omega$-deformed $\bR^2$ \cite{Shadchin:2006yz} dressed by the KK modes on $S^1$, much like the 5d instanton partition function of \cite{Nekrasov:2003rj} on $\bR^4_{\epsilon_1,\epsilon_2} \times S^1$). The precise identification of parameters depends on the geometry.

We expect the same method to work on other three-manifolds, for instance for the lens space index on $S^3_b/\bZ_p$ \cite{Benini:2011nc, Alday:2012au}, and also in four dimensions on manifolds like $S^3 \times S^1$ \cite{Romelsberger:2005eg, Gadde:2009kb}, $S^3_b/\bZ_p \times S^1$ \cite{Benini:2011nc, Razamat:2013opa} and $S^2 \times T^2$ \cite{Closset:2013sxa}. We leave this for future work.

The paper is organized as follows. In section \ref{sec: S3b} we study the case of $S^3_b$: we analyze the BPS equations and their solutions, we study the effect of the new deformation term responsible for Higgs branch localization, and write the general form of the partition function. We conclude with the example of a $U(N)$ gauge theory with (anti)fundamentals \cite{Taki:2013opa}. In section \ref{sec: S2xS1} we do the same in the case of $S^2 \times S^1$. We also consider the example of $U(N)$ \cite{Hwang:2012jh}, and show that $S^3_b$ and $S^2\times S^1$ are controlled by the very same vortex partition function.

\

\emph{Note added.} When this work was under completion, we became aware of \cite{Fujitsuka:2013fga} which has substantial overlap with our paper.

\section{Higgs branch localization on $S^3_b$}
\label{sec: S3b}

We start by studying the path integral of three-dimensional $\cN=2$ R-symmetric Yang-Mills-Chern-Simons-matter theories on the squashed three-sphere $S^3_b$, where $b$ is a squashing parameter, and its supersymmetric localization. Such a path integral has been computed, with localization techniques, in \cite{Hama:2011ea}, building on the works \cite{Kapustin:2009kz, Jafferis:2010un, Hama:2010av} (see also \cite{Imamura:2011wg}). In their framework the path integral is dominated by BPS configurations that look like a classical Coulomb branch: the only non-vanishing field is an adjoint-valued real scalar in the vector multiplet (together with an auxiliary scalar), which can be diagonalized to the maximal torus. We thus dub this ``Coulomb branch localization'': the resulting expression in \cite{Hama:2011ea} is a matrix-model-like partition function, that we review in section \ref{sec: S3 Coulomb branch}.

Our goal is to perform localization in a different way, by including an extra $\cQ$-exact term in the deformation action,%
\footnote{$\cQ$ is a supercharge, and the path integral is not affected by the insertion of $\cQ$-exact terms \cite{Witten:1988ze, Witten:1991zz}.}
so that the path integral is dominated by BPS configurations that look like vortex strings at a northern circle and antivortex strings at a southern circle. Vortices exist on the Higgs branch, therefore we dub this \emph{Higgs branch localization}, as in \cite{Benini:2012ui}.

We will focus on a special class of backgrounds with three-sphere topology, the squashed three-sphere $S^3_b$ of \cite{Hama:2011ea} as we said, because our goal is to spell out how Higgs branch localization works. Much more general backgrounds are possible on $S^3$ \cite{Klare:2012gn, Closset:2012ru}, and we expect Higgs branch localization to be extendable to all those backgrounds easily. Moreover it has been shown in \cite{Closset:2013vra} that the supersymmetric partition function depends on the background through a single continuous parameter $b$ (there might be multiple connected components, though), therefore the computation on $S^3_b$ produces the full set of possible functions one can obtain in this way from the field theory.

\subsection{Killing spinors on $S^3_b$}\label{sec: Killing spinors on S3b}

We consider a squashed three-sphere $S^3_b$ with metric \cite{Hama:2011ea}
\be
\label{metricS3b}
ds^2 = f(\theta)^2 d\theta^2 + \tilde\ell^2 \sin^2 \theta\, d\chi^2  + \ell^2 \cos^2 \theta\, d\varphi^2 \;,
\ee
where $f(\theta) = \sqrt{\ell^2 \sin^2 \theta + \tilde\ell^2 \cos^2 \theta}$ and the squashing parameter $b$ is defined as $b=\sqrt{\tilde\ell/\ell}$. The ranges of coordinates are $\theta \in [0,\frac\pi2]$ and $\chi,\varphi \in [0,2\pi)$. In fact, as apparent in \cite{Hama:2011ea} and remarked in \cite{Martelli:2011fu} (see also \cite{Alday:2013lba}), any function $f(\theta)$ which asymptotes to $\tilde\ell, \ell$ at $\theta = 0, \frac\pi2$ respectively and which gives a smooth metric, would lead to the same results. We choose the vielbein one-forms as
\be
e^{\underline1} = \ell \cos\theta\, d\varphi \;,\qquad e^{\underline2} = -\tilde\ell \sin\theta\, d\chi \;,\qquad e^{\underline3} = f(\theta) d\theta \;,
\ee
yielding the non-zero components of the spin connection  $\omega^{\underline{13}} = -\frac\ell f\sin\theta\, d\varphi$ and $\omega^{\underline{23}} = -\frac{\tilde\ell}f \cos\theta d\chi$. We underline the flat coordinates in this frame. We also turn on a background gauge field that couples to the $U(1)_R$ R-symmetry current:
\be \label{Rbgf}
V = \frac12 \Big( 1 - \frac\ell{f} \Big) \, d\varphi + \frac12 \Big( 1 - \frac{\tilde\ell}{f} \Big)\, d\chi \;.
\ee
The twisted Killing spinor equation%
\footnote{In our conventions $D_\mu = \partial_\mu + \frac14 \omega_\mu^{ab} \gamma_{ab} - i V_\mu$. Charge conjugation is $\epsilon^c = C\epsilon^* = \gamma_{\underline2}\epsilon^*$, having chosen $C=\gamma_{\underline2}$.}
$D_\mu \epsilon = \gamma_\mu \hat\epsilon$ (where $\gamma_{\underline a}$ are Pauli matrices) is then solved by the two spinors \cite{Hama:2011ea}
\be
\epsilon = \frac1{\sqrt2} \mat{ e^{-\frac i2 (\varphi + \chi - \theta)} \\ -e^{- \frac i2 (\varphi + \chi + \theta)} } \;,\qquad\qquad
\bar\epsilon = \frac1{\sqrt2} \mat{ e^{\frac i2(\varphi + \chi + \theta)} \\ e^{\frac i2(\varphi + \chi - \theta)} }
\ee
by assigning R-charges $R[\epsilon]=-1$ and $R[\bar\epsilon]=1$. In fact they satisfy
\be
D_\mu \epsilon = \frac i{2f} \gamma_\mu \epsilon \;,\qquad\qquad\qquad D_\mu \bar\epsilon = \frac i{2f} \gamma_\mu \bar\epsilon \;.
\ee
We also define the charge conjugate spinor $\tilde\epsilon \equiv - \bar\epsilon^c = i\epsilon$. For spinor conventions see appendix \ref{app: spionor conventions}.

Two bilinears that we will need are:
\be
\xi^{\underline a} = i \bar\epsilon \gamma^{\underline a} \epsilon = - \epsilon^\dag \gamma^{\underline a} \epsilon = \big( -i\cos \theta,  i\sin\theta , 0 \big) \;,\qquad\qquad
\bar\epsilon \epsilon = i \epsilon^\dag\epsilon = i \;.
\ee
Using the coordinate frame $(\varphi,\chi,\theta)$ we have
\be
\xi^\mu = i \bar\epsilon \gamma^\mu \epsilon = \Big( \frac1\ell , \frac1{\tilde\ell}, 0 \Big) \;.
\ee
There are also two useful scalar bilinears, $\rho$ and $\alpha$ defined in (\ref{def of bilinears}), which take values $\rho = 0$ and $\alpha = -\frac{1}{f} -\xi^\mu V_\mu = - \frac12 \big( \frac1\ell + \frac1{\tilde\ell}\big)$. Therefore the commutator of SUSY transformations (\ref{SUSY commutators}) is
\be
[\delta_\epsilon, \delta_{\bar\epsilon}] = \cL^A_\xi  -\sigma -  \frac i2 \left(\frac1\ell + \frac1{\tilde\ell} \right) R \;.
\ee

It will be useful to perform a frame rotation such that the Killing vector field $\xi = \xi^\mu \partial_\mu$ becomes one of the frame vectors. We then define the non-underlined frame and its dual basis of vectors:
\bea
e^1 &= -f(\theta) d\theta \qquad& e^2 &= \cos\theta\sin\theta\, (\ell\, d\varphi - \tilde\ell \, d\chi) \qquad& e^3 &= \ell \cos^2\theta\, d\varphi + \tilde\ell \sin^2\theta\, d\chi \\
e_1 &= -f(\theta)^{-1} \partial_\theta \qquad& e_2 &= \ell^{-1} \tan\theta\, \partial_\varphi - \tilde\ell^{-1} \cot\theta\, \partial_\chi \qquad& e_3 &= \ell^{-1} \partial_\varphi + \tilde\ell^{-1} \partial_\chi \;.
\eea
In particular $\xi = e_3$. In this basis the spin connection reads
\be
\omega^{ab} = \mat{
0 &  -\frac\ell f \sin^2\theta\, d\varphi - \frac{\tilde\ell}f \cos^2\theta\, d\chi  &
\frac{\sin2\theta}{2f} (-\ell\, d\varphi + \tilde\ell\, d\chi) \\
\frac\ell f \sin^2\theta\, d\varphi + \frac{\tilde\ell}f \cos^2\theta\, d\chi  & 0 & -d\theta\\
\frac{\sin2\theta}{2f} (\ell\, d\varphi - \tilde\ell\, d\chi) & d\theta & 0 }
\ee
and the Killing spinors become
\be
\label{kss3b}
\epsilon = \mat{ 0 \\ -e^{-\frac i2(\varphi + \chi)} } \;,\qquad\qquad
\bar\epsilon = \mat{ e^{\frac i2(\varphi + \chi )} \\ 0 }
\ee
as well as $\tilde\epsilon = -\bar\epsilon^c = i\epsilon$. The relation between the two bases is $e^a = \Big( \begin{smallmatrix} 0 & 0 & -1 \\ \sin\theta & \cos\theta & 0 \\ \cos\theta & -\sin\theta & 0 \end{smallmatrix} \Big)_{a\underline a} e^{\underline a}$, where the matrix has determinant one. In the rest of this section we will use the non-underlined frame.

To conclude let us describe the metric of the squashed three-sphere using Hopf coordinates $\phi_H = \varphi - \chi$ and $\psi_H = \varphi + \chi$, in which the Killing vector $\xi = \big( \frac1\ell + \frac1{\tilde\ell} \big) \partial_{\psi_H} + \big(\frac1\ell - \frac1{\tilde\ell} \big)\partial_{\phi_H}$. On the round sphere of radius 1, $\xi = 2\partial_{\psi_H}$ generates pure motion around the Hopf fiber, whilst the squashing introduces an additional rotation of the base space $S^2$ with fixed points at $\theta=0$ and $\theta = \frac\pi2$. The metric \eqref{metricS3b} reads in these coordinates:
\begin{multline}
\label{Hopf coordinates}
ds^2 = f(\theta)^2 d\theta^2 + \frac{\ell^2 \tilde\ell^2 \sin^2 2\theta}{4(\ell^2\cos^2\theta + \tilde\ell^2 \sin^2\theta)} d\phi_H^2 \\
+ \frac14(\ell^2\cos^2\theta + \tilde\ell^2\sin^2\theta) \Big( d\psi_H + \frac{\ell^2 \cos^2\theta - \tilde\ell^2 \sin^2\theta}{\ell^2 \cos^2\theta + \tilde\ell^2 \sin^2\theta} d\phi_H \Big)^2 \;.
\end{multline}
In fact one could instead take $\partial_{\phi_H}$ as the Hopf vector field, and rewrite the metric in the same form as above but with $\psi_H \leftrightarrow \phi_H$.

\subsection{The BPS equations}

We will now consider the BPS equations for vector and chiral multiplets, and how they can be obtained as the zero-locus of the bosonic part of a $\cQ$-exact deformation action. See appendix \ref{app: SUSY and actions} for the SUSY transformations.

First we define
\be
\label{def Ws}
W^r = \frac12 \varepsilon^{rmn} F_{mn} \;,\qquad\qquad F_{mn} = \varepsilon_{mnr} W^r \;,
\ee
so that $\frac12 F_{mn} F^{mn} = W_m W^m$. Then, from \eqref{gaugemultiplet}, the BPS equations for the vector multiplet are
\bea
\label{BPS eqns vector}
0 &= \cQ\lambda = i \big( W_\mu + D_\mu\sigma \big) \gamma^\mu \epsilon - \Big( D + \frac\sigma f \Big)\epsilon \\
0 &= \cQ \lambda^\dag = -i \tilde\epsilon^\dag \gamma^\mu \big( W_\mu - D_\mu\sigma \big) + \tilde\epsilon^\dag \Big( D+\frac\sigma f \Big) \;.
\eea
Recall that in Euclidean signature we regard $\lambda$ and $\lambda^\dag$ as independent fields.
It is convenient to use the non-underlined frame and the Killing spinors in \eqref{kss3b}; after taking sums and differences of the components, we get the BPS equations:
\be
\label{BPS eqns YM non-def}
0 = W_1 -i D_2\sigma \;,\qquad 0= W_2 + i D_1\sigma \;,\qquad 0 = W_3 - i \Big( D + \frac\sigma f \Big) \;,\qquad 0 = D_3\sigma \;.
\ee

In fact---as it is standard---the equations (\ref{BPS eqns YM non-def}) can be derived as the zero-locus of the bosonic part of a $\cQ$-exact deformation action, whose Lagrangian is
\be
\label{YM def action}
\cL^\text{def}_\text{YM} = \cQ \Tr \Big[ \frac{(\cQ\lambda)^\ddagger \lambda + \lambda^\dag (\cQ \lambda^\dag )^\ddagger}{4} \Big] \;.
\ee
Here the action of the formal adjoint operator $\ddagger$ on $\cQ\lambda$  and $\cQ\lambda^\dag$ is:
\bea
(\cQ\lambda)^\ddagger &= \epsilon^\dag \Big[ - i \gamma^\mu (W_\mu + D_\mu \sigma^\dag) - \Big( D + \frac{\sigma^\dag}f \Big) \Big] \\
(\cQ\lambda^\dag)^\ddagger &= \Big[ i (W_\mu - D_\mu\sigma^\dag) \gamma^\mu + \Big( D + \frac{\sigma^\dag}f \Big) \Big] \tilde\epsilon \;,
\eea
where we treat $\sigma$ as a complex field. The operator $\ddagger$ reduces to $\dag$ when $A_\mu$ and $D$ are taken real. Decomposing $\sigma = \sigma_R + i\sigma_I$ into its real and imaginary parts, we find that the bosonic part of $\cL^\text{def}_\text{YM}$ is a positive sum of squares:
\begin{multline}
\label{gmlocals3}
\frac14 \Tr \Big[ (\cQ\lambda)^\ddagger \cQ\lambda + \cQ\lambda^\dag (\cQ\lambda^\dag)^\ddagger \Big] = \Tr \bigg\{ \frac12 \big( W_1+ D _2 \sigma_I \big)^2 + \frac12 \big( W_2 - D_1\sigma_I)^2 \\
+ \frac12 \Big(W_3 + \frac{\sigma_I}f \Big)^2 + \frac12 ( D_3\sigma_I )^2+ \frac12 \sum_{a=1,2,3} (D_a\sigma_R)^2 + \frac12 \Big( D + \frac{\sigma_R}f \Big)^2 \bigg\} \;.
\end{multline}
If we restrict to real fields, $\sigma_I = 0$, from the zero locus of this action we recover the localization locus $F_{\mu\nu} = 0$ and $\sigma = - fD = \text{const}$, as in \cite{Kapustin:2009kz}. On a three-sphere, $F_{\mu\nu} = 0$ allows us to set $A_\mu =0$, then $D_\mu\sigma = \partial_\mu\sigma$ and finally $\sigma$ can be diagonalized. On the other hand the equations (\ref{BPS eqns YM non-def}) allow for more general solutions with complex $\sigma$.

As in \cite{Benini:2012ui}, Higgs branch localization can be achieved by adding another $\cQ$-exact term to the deformation action. Consider
\be
\label{def action H}
\cL^\text{def}_\text{H} = \cQ \Tr \Big[ \frac{i (\epsilon^\dag \lambda - \lambda^\dag \tilde\epsilon) \, H(\phi)}2 \Big] \;,
\ee
whose bosonic part is
\be
\label{hlocals3}
\cL^\text{def}_\text{H} \Big|_\text{bos} = \Tr \Big[ \Big( W_3 - i \big( D + \tfrac\sigma f \big) \Big) H(\phi) \Big] \;.
\ee
The action $S^\text{def}_\text{H} = \int \cL^\text{def}_\text{H}$ is both $\cQ$-exact and $\cQ$-closed.%
\footnote{While exactness is manifest in (\ref{def action H}), closeness follows from an argument in \cite{Hosomichi:2012ek}. If $\epsilon^\dag, \tilde\epsilon$ were fields, the integral of the trace in  (\ref{def action H}) would be invariant under $\cQ^2$ because it is a neutral scalar. Therefore if $\epsilon$ is invariant under the bosonic operator $\cQ^2$, then $S^\text{def}_\text{H}$ is $\cQ$-closed. It is easy to check that $\cQ^2 \epsilon = \cL_\xi \epsilon + \frac i2 (\ell^{-1} + \tilde\ell^{-1}) \epsilon = 0$.}
$H(\phi)$ is a generic real function of the complex scalar fields $\phi,\phi^\dag$ in chiral multiplets,%
\footnote{If we want to be sure that $\cL^\text{def}_\text{H}$ does not change the vacuum structure of the theory, we should limit ourselves to functions $H$ that do not modify the behavior of the action at infinity in field space \cite{Witten:1992xu}. This is the case if $H(\phi)$ is quadratic.}
taking values in the adjoint representation.
Actually one could even consider more general functions $H(\phi,\sigma)$---and we mention the interesting fact that $H(\phi,\sigma) = H(\phi) + \kappa \sigma_I$ would lead to Yang-Mills-Chern-Simons vortex equations---but we will not do so in this paper.

The bosonic part of the new deformation term $\cL^\text{def}_\text{H}$ is not positive definite. However if we consider the sum $\cL^\text{def}_\text{YM} + \cL^\text{def}_\text{H}$, the auxiliary field $D$ appears quadratically without derivatives and can be integrated out exactly by performing the Gaussian path integral. This corresponds to imposing
\be
\label{D-term equation}
D + \frac{\sigma_R}f = i H(\phi) \;,
\ee
in other words $D + \sigma_R/f$ is formally taken out of the real contour. The bosonic part of what we are left with is a positive sum of squares:
\begin{multline}
\cL^\text{def}_\text{YM} + \cL^\text{def}_\text{H} \Big|_\text{$D$, bos} = \Tr \bigg[ \frac12 \big( W_1 + D_2\sigma_I \big)^2 + \frac12 \big( W_2 - D_1\sigma_I \big)^2 \\
+ \frac12 \Big( W_3 + \frac{\sigma_I}f + H(\phi) \Big)^2 + \frac12 (D_3\sigma_I)^2 + \frac12 \sum_{a=1,2,3} (D_a\sigma_R)^2 \bigg] \;.
\end{multline}
The BPS equations describing its zero-locus are then
\be
\label{BPS eqns YM def}
0 = W_1 + D_2\sigma_I \;,\quad 0= W_2 - D_1\sigma_I \;,\quad 0 = W_3 + \frac{\sigma_I}f + H(\phi) \;,\quad 0 = D_3\sigma_I \;,\quad 0= D_a\sigma_R \;. \;
\ee
These equations differ from (\ref{BPS eqns YM non-def}) only by the fact that the ``D-term equation'' (\ref{D-term equation}) has been imposed.

Let us now consider the chiral multiplets, transforming in some (possibly reducible) representation of the gauge and flavor symmetry group. At this point it is useful to introduce some notation. We call $\fR$ the (possibly reducible) representation of the gauge and flavor symmetry group under which all chiral multiplets transform. Accordingly, we consider a vector multiplet for the full gauge and flavor symmetry, the components for the gauge group being dynamical and those for the flavor group being external, and whose real scalar we call $\fS$. On a supersymmetric background, external vector multiplets should satisfy the same BPS equations (\ref{BPS eqns vector}), but of course they do not have a kinetic action. Real expectation values of the external components of $\fS$ are the so-called real masses, so coupling a chiral multiplet in representation $\fR$ to $\fS$ includes real masses as well.%
\footnote{In our discussion we are not completely general. In three dimensions, the flavor symmetry group usually includes topological (or magnetic) symmetries which do not act on the microscopic chiral multiplets in the Lagrangian, but rather on monopole operators, and real mass parameters can be included for those symmetries as well. For instance, a $U(1)$ gauge theory has a $U(1)_T$ topological symmetry and a real mass for it is the Fayet-Iliopoulos term. However in our formalism FI terms have to be included by hand, rather than turning on the corresponding component of $\fS$.}
On the other hand, we decompose $\fR$ into irreducible representations of the gauge group: $\fR = \bigoplus_i \cR_i$. In this notation, each chiral multiplet in representation $\cR_i$ couples to $\sigma$ and to its real mass term $m_i$. The projection of $\fS$ on the representation $\cR_i$ is $\fS \big|_{\cR_i} = \sigma + m_i$.

For each irreducible gauge representation $\cR$, the BPS equations $\cQ\psi = \cQ\psi^\dag = 0$ give
\bea
0 &= D_3\phi - \Big( \sigma + m + i\frac qf \Big) \phi \qquad\qquad &
0 &= e^{-\frac i2 (\chi + \varphi)} (D_1 - iD_2)\phi + ie^{\frac i2 (\chi+\varphi)} F \\
0 &= D_3 \phi^\dag + \phi^\dag \Big( \sigma + m + i\frac qf \Big) &
0 &= e^{\frac i2 (\chi+\varphi)}(D_1 + iD_2) \phi^\dag + ie^{-\frac i2 (\chi + \varphi)} F^\dag \;,
\eea
where $m$ is the mass and $q$ is the R-charge (all fields in $\cR$ must have the same mass and R-charge). Imposing the reality conditions $\phi = (\phi^\dag)^\dag$, $F = (F^\dag)^\dag$ and decomposing $\sigma$ into real and imaginary parts as before, the equations simplify to
\be
\label{BPS eqns mat}
( \sigma_R + m ) \phi = 0 \;,\qquad D_3\phi - i \Big(\sigma_I + \frac qf \Big)\phi = 0 \;,\qquad (D_1 - iD_2)\phi = 0 \;,\qquad F = 0 \;.
\ee
In passing we note that, since $\xi = e_3$ and using the first equation $\re\fS\big|_{\cR_i} \phi = 0$, the second one is
\be
0 = \xi^\mu\big( \partial_\mu - iA_\mu - iqV_\mu\big)\phi - i \Big(\sigma_I + \frac qf \Big)\phi = \Big[ \cL_\xi^A - \frac{iq}2 \Big( \frac1\ell + \frac1{\tilde\ell} \Big) - \fS \big|_{\cR_i} \Big]\phi = \cQ^2 \phi \;.
\ee

As before, these equations can also be obtained from the canonical deformation action
\be
\label{mat def action}
\cL^\text{def}_\text{mat} = \cQ \, \frac{(\cQ\psi)^\dagger \psi + \psi^\dagger(\cQ\psi^\dagger)^\dagger}4 \;.
\ee
Up to total derivatives, its bosonic part reads
\be
\cL^\text{def}_\text{mat} \Big|_{\text{bos}} =\frac12 \Big| D_3\phi - i \big(\sigma_I + \tfrac qf \big) \phi \Big|^2 + \frac12 \big| (D_1 - iD_2)\phi \big|^2 + \frac12 \big| (\sigma_R + m) \phi \big|^2 + \frac12 |F|^2 \;,
\ee
where we recognize once again the BPS equations.

To conclude this section, let us rewrite the BPS equations in components since it will be useful later on. For the vector multiplet we find
\bea
\label{BPS eqns components YM}
\ell^{-1} \tilde\ell^{-1} F_{\varphi\chi} &= \big( - \ell^{-1} \sin^2\theta\, D_\varphi + \tilde\ell^{-1} \cos^2\theta\, D_\chi \big) \sigma_I \hspace{-.5cm} \\
\ell^{-1} F_{\theta\varphi} + \tilde\ell^{-1} F_{\theta\chi} &= - D_\theta \sigma_I &
0 &= \big( \ell^{-1} D_\varphi + \tilde\ell^{-1} D_\chi \big) \sigma_I \\
\ell^{-1} \tan\theta F_{\theta\varphi} - \tilde\ell^{-1} \cot\theta\, F_{\theta\chi} &= f(\theta)\, H(\phi) + \sigma_I &
0  &= D_\mu \sigma_R \;,
\eea
and for the chiral multiplet we get $0 = (\sigma_R + m) \phi = F$ as well as
\bea
\label{BPS eqns components matter}
\big( \ell^{-1} D_\varphi + \tilde\ell^{-1} D_\chi \big)\phi &= i \Big( \sigma_I + \frac qf \Big) \phi \\
\big( f(\theta)^{-1} D_\theta + i \ell^{-1} \tan\theta\, D_\varphi - i \tilde\ell^{-1} \cot\theta\, D_\chi \big) \phi &= 0 \;.
\eea

\subsection{BPS solutions: Coulomb, Higgs and vortices}
\label{sec: solutionsS3}

We will now analyze the solutions to (\ref{BPS eqns YM non-def}), (\ref{BPS eqns YM def}) and (\ref{BPS eqns mat}). First, let us recall the solutions for the standard choice $H(\phi) = 0$.

\paragraph{Coulomb-like solutions.}
Consider (\ref{BPS eqns YM non-def}) and (\ref{BPS eqns mat}). We solve them along a ``real'' contour where $A_\mu, \sigma, D$ are real, in particular $\sigma_I = 0$, and $(\phi,\phi^\dag)$, $(F,F^\dag)$ are conjugate pairs. Moreover we assume that all chiral multiplets have positive R-charge. As mentioned before, the solutions are \cite{Kapustin:2009kz}
\be
A_\mu =0 \;,\qquad\qquad \sigma = -fD = \text{const} \;,\qquad\qquad \phi = F = 0 \;.
\ee
Let us check that there are no solutions with non-trivial $\phi$. We can Fourier expand along the compact directions $\varphi,\chi$:
\be
\phi(\theta,\varphi,\chi) = \sum_{m,n \,\in\, \bZ} c_{mn}(\theta) \, e^{i n\varphi} e^{im\chi} \;.
\ee
The first equation in (\ref{BPS eqns components matter}) imposes the constraint $q = 2(m\ell + n\tilde\ell)/(\ell + \tilde\ell)$ for $m,n\in \bZ$.
In particular for incommensurable values of $\ell,\tilde\ell$, either $q$ is one of the special values above and in this case there is only one Fourier mode $(m,n)$, or $\phi=0$ is the only solution. Assuming that $\ell, \tilde\ell$ are incommensurable and that $m,n$ are fixed and solve the constraint, the second equation in (\ref{BPS eqns components matter}) reduces to $\big( \sin2\theta\, \partial_\theta + q\cos2\theta + L f(\theta) \big) \phi = 0$ with $L = 2(m-n)/(\ell + \tilde\ell)$. The solution is
\be
\phi(\theta,\varphi,\chi) = \Big( \frac{1-s(\theta)}{1+s(\theta)} \Big)^{\tfrac{L \ell}4} \Big( \frac{1-\tilde s(\theta)}{1+\tilde s(\theta)} \Big)^{-\tfrac{L \tilde\ell}4} (\sin2\theta)^{-q/2} \, e^{i n\varphi} e^{im\chi}
\ee
with
\be
s(\theta) = \sqrt{ \frac{\ell^2 + \tilde\ell^2 - (\ell^2 - \tilde\ell^2)\cos2\theta}{2\ell^2}} \;,\qquad\qquad \tilde s(\theta) = \sqrt{ \frac{\ell^2 + \tilde\ell^2 - (\ell^2 - \tilde\ell^2)\cos2\theta}{2\tilde\ell^2}} \;.
\ee
The functions $s,\tilde s$ are monotonic and positive, with $s(0) = \tilde s(\frac\pi2)^{-1} = \tilde\ell/\ell$ and $s(\frac\pi2) = \tilde s(0) = 1$.
For $q>0$ there are no smooth solutions. For $q=0$ (then $m=n=0$) there is the constant Higgs-like solution $\phi = \phi_0$ that we will re-encounter below (in this case, $\sigma_R$ is constrained by $(\sigma_R+m)\phi = 0$), but we will not consider it here since we assumed that R-charges are positive.

\

Now let us study the new solutions with non-trivial $H(\phi)$. We integrate $D$ out first, \ie{} we solve (\ref{BPS eqns YM def}) and (\ref{BPS eqns mat}) and impose a ``real'' contour for all fields but $D$ (in particular $\sigma_I = 0$ again). We also take vanishing R-charges, $q=0$: arbitrary R-charges can be recovered by analytic continuation of the final result in the real masses, as in \cite{Benini:2012ui}. We make the following choice for $H(\phi)$:
\be
\label{H function}
H(\phi) = \zeta - \sum_{i, a} T^a_\text{adj} \; \phi_i^\dag T^a_{\cR_i} \phi_i
\ee
where the sum is over the representations $\cR_i$ and the gauge symmetry generators $T^a$ in representation $\cR_i$. The adjoint-valued parameter $\zeta$ is defined as
\be
\label{expansion fake FI}
\zeta = \sum_{a:\, U(1)} \zeta_a h_a \;,
\ee
\ie{} a sum over the Cartan generators $h_a$ of the Abelian factors in the gauge group, in terms of the real parameters $\zeta_a$. We find the following classes of solutions.

\paragraph{Deformed Coulomb branch.} It is characterized by $\phi = 0$, therefore from (\ref{BPS eqns YM def}):
\be
F = \zeta \sin\theta \cos\theta\, f(\theta)\, d\theta \wedge (\ell\, d\varphi - \tilde\ell \, d\chi) \;.
\ee
Since $S^3_b$ has trivial second cohomology, any line bundle is trivial and we can find a globally defined and smooth potential:\be \label{deformedgaugefield}
A = \zeta \Big[ \big( G(\theta) - G(\pi/2) \big) \ell\, d\varphi + \big( G(0) - G(\theta) \big) \tilde\ell \, d\chi \Big]
\ee
where $G'(\theta) = \sin\theta \cos\theta\, f(\theta)$. We find
\be
G(\theta) = \frac{\big( \ell^2 + \tilde\ell^2 - (\ell^2 - \tilde\ell^2) \cos2\theta \big)^{3/2}}{6\sqrt2\, (\ell^2 - \tilde\ell^2)} + \text{const} \;,\quad G\big(\tfrac\pi2\big) - G(0) = \frac{\ell^2 + \ell\tilde\ell + \tilde\ell^2}{3(\ell + \tilde\ell)} = \frac{\vol(S^3_b)}{4 \pi^2 \ell \tilde \ell} \;.
\ee
The scalar $\sigma$ is constant and it commutes with $F$, in particular we can choose a gauge where it is along the Cartan subalgebra.

\paragraph{Higgs-like solutions.} They are characterized by $H(\phi) = 0$ (we will relax this condition momentarily). This implies $F_{\mu\nu} = 0$ and, choosing $A_\mu = 0$, also $0 = \partial_\mu \sigma = \partial_\mu \phi$ (one has to exclude non-constant solutions for $\phi$ with the same argument as above). Therefore $\sigma$ can be diagonalized, and one is left with the algebraic equations
\be
\label{Higgs branch equations}
H(\phi) = 0 \;,\qquad\qquad \big( \sigma + m_i \big) \, \phi_i = 0 \qquad\forall\, i \;.
\ee
The last equation can be more compactly written as $\fS \phi = 0$. These are the standard D-term equations, and their solutions strongly depend on the gauge group and matter content of the theory.

We will be interested in gauge groups and matter representations for which generic parameters $\zeta_a$ and generic masses $m_i$ lead to solutions to (\ref{Higgs branch equations}) that completely break the gauge group. More specifically, we will be focusing on theories for which the Coulomb branch parameters $\sigma_\alpha$, for $\alpha = 1,\ldots,\rank G$, are fixed (depending on the Higgs-like solution) in terms of the masses $m_i$, and for generic masses they are different breaking the gauge group to $U(1)^{\rank G}$. Each $U(1)$ is then Higgsed by one component of $\phi$, along a weight $w\in\fR$, getting VEV. One gets a discrete set of Higgs vacua. If the gauge group is not completely broken (including the case of an unbroken discrete gauge group), or if some continuous Higgs branch is left, the situation is more involved and we will not study it here.

\paragraph{Vortices.} Each Higgs-like solution is accompanied by a tower of other solutions with arbitrary numbers of vortices at the north and at the south circles (the Higgs-like solution should be thought of as the one with zero vortex numbers). To see this, expand the BPS equations around $\theta=0$ at first order in $\theta$. Defining the coordinate $r = \tilde\ell \theta$, the metric reads
\be
ds^2 \,\simeq\, dr^2 + r^2 d\chi^2 + \ell^2 d\varphi^2 \qquad\qquad\text{around $\theta=0$}
\ee
which is $\bR^2 \times S^1$. The BPS equations (\ref{BPS eqns components YM}) and (\ref{BPS eqns components matter}) reduce to
\bea
\label{vortexeqnN}
r^{-1} F_{r\chi} &= - H(\phi) \qquad\qquad & F_{r\varphi} &= - \frac\ell{\tilde\ell} \, F_{r\chi} \qquad\qquad & F_{\varphi\chi} &= 0 \\
0 &= \Big( D_r - \frac ir D_\chi \Big) \phi \qquad\qquad & D_\varphi \phi &= - \frac\ell{\tilde\ell} D_\chi\phi \;.
\eea
The two equations on the left are the usual vortex equations%
\footnote{They are more conventionally antivortex equations, the difference being only the orientation.}
on $\bR^2$, while the other equations complete the solutions to vortices on $\bR^2 \times S^1$ once the solutions on $\bR^2$ are found. The equations cannot be solved analytically, therefore let us qualitatively describe the solutions in the $U(1)$ case with a single chiral of charge 1, since---up to a rescaling of the charge---this is the generic situation once the gauge group has been broken to $U(1)^{\rank G}$ by the VEV of $\sigma$. We take $\zeta > 0$, in order to have solutions. Far from the core of the vortex, for $r \gg \sqrt{m/\zeta}$ (the integer $m$ will be defined momentarily), we have $0 = H(\phi) = F_{r\chi} = F_{r\varphi}$ therefore
\be
\label{far from core}
\phi \,\simeq\, \sqrt\zeta\, e^{-in\varphi - im\chi} \;,\qquad\qquad A \,\simeq\, -n\, d\varphi - m \, d\chi \;.
\ee
Stokes' theorem on $\bR^2$ implies $\frac1{2\pi} \int F = - m$, \ie{} $m$ is the vortex number at the north circle (while $n$ will be interpreted below). At the core of the vortex $\phi$ has to vanish in order to be smooth (if $m\neq 0$), therefore close to the core
\be
\label{close to core}
\begin{aligned}
\phi &\,\simeq\, B (r e^{-i\chi})^m e^{-in\varphi} \;,\qquad\quad F \,\simeq\, \zeta r \, dr \wedge \Big( \frac\ell{\tilde\ell} \, d\varphi - d\chi \Big) \\
A &\,\simeq\, \Big( -n - \frac\ell{\tilde\ell} \, m + \zeta \, \frac\ell{\tilde\ell} \, \frac{r^2}2 \Big) d\varphi - \zeta \, \frac{r^2}2 \, d\chi
\end{aligned}
\qquad\qquad \text{for } r \ll \sqrt{m/\zeta}
\ee
where $B$ is some constant. In particular, smoothness of $\phi$ requires $m \in \bZ_{\geq 0}$. Note that $\phi$ vanishes only at $r=0$, therefore
\be
\tilde\ell\, A_\varphi + \ell\, A_\chi = - \tilde\ell\, n - \ell\, m
\ee
holds exactly. If we approximate $r^{-1} F_{r\chi}$ by a step function on a disk times $-\zeta$, we get that the size of the vortex is of order $\sqrt{m/\zeta}$ justifying the limits we took. In the limit $\zeta \to \infty$ the vortices squeeze to zero-size, therefore the first-order approximation of the equations around $\theta = 0$ is consistent.

We can similarly study the BPS equations expanded around $\theta = \frac\pi2$ at first order in $\frac\pi2-\theta$, defining a coordinate $\tilde r = \ell\left( \frac\pi2-\theta\right)$. As before, the equations reduce to the 2d antivortex equations (as the orientation induced from $S^3_b$ is opposite) besides some other equations that complete the solutions to 3d.
For a $U(1)$ gauge theory with a single chiral, the analysis above goes through {\it mutatis mutandis}. Far from the core of the vortex, for $\tilde r \gg \sqrt{n/\zeta}$, we have the same asymptotic behavior as in (\ref{far from core}). Stokes' theorem on $\bR^2$ implies $\frac1{2\pi} \int F = - n$, \ie{} $n$ is the antivortex number at the south circle, and the analysis of the solution for $\tilde r \ll \sqrt{n/\zeta}$ reveals that $n \in \bZ_{\geq0}$. The behavior of the fields (\ref{far from core}) in the intermediate region, far from both cores, provides a link of parameters between the two cores and it is indeed a solution of the full BPS equations.

For finite values of $\zeta$, both curvature and finite size effects play a r\^ole. From the second and third equations on the left in \eqref{BPS eqns components YM}, integrating over the sphere one can obtain
\be
-4\pi^2 \ell \int F_{\theta \chi} \, d\theta = 4\pi^2 \tilde\ell \int F_{\theta \varphi} \, d\theta = \int H(\phi) \, \dvol_{S^3_b} \;\leq\; \zeta \vol(S^3_b) \;,
\ee
where we used that $H(\phi)$ is bounded by $0 \leq H(\phi) \leq \zeta$ on vortex solutions, and vortex solutions have only $\theta$ dependence. Still working in a gauge with smooth and globally defined connection $A$, we can define the vortex numbers $m,n$ at the north and south circle as the winding numbers of $\phi$ around $\chi,\varphi$ respectively. The analyses at the cores are still valid, therefore $m,n\in \bZ_\geq0$ and
\be
- \frac{A_\varphi(0)}\ell = - \frac{A_\chi\big( \frac\pi2\big)}{\tilde\ell} = \frac n\ell + \frac m{\tilde\ell} \;.
\ee
Then the bound above implies a bound on the vortex and antivortex numbers:
\be
\label{bound}
b\,n + b^{-1}m  \;\leq\;  \zeta \, \frac{\vol(S^3_b)}{4\pi^2 \sqrt{\ell \tilde\ell}} \;.
\ee
We conclude that for finite values of $\zeta$ there is a finite number of vortex/antivortex solutions on the squashed three-sphere; when the bound is saturated, the chiral field $\phi$ actually vanishes and the gauge field is as in the deformed Coulomb branch described before. We thus get a nice picture of the structure of solutions as we continuously increase $\zeta$ from $0$ to $+\infty$. The Coulomb branch solution is continuously deformed into the deformed Coulomb branch solution; as $\zeta$ crosses one of the thresholds, proportional to $bn + b^{-1}m$, a new (anti)vortex solution branches out, in which the value of the matter field is infinitesimal at the threshold and increases further on. This picture will be useful in the next section to understand how localization changes as we change $\zeta$ continuously.

\

For gauge groups of rank larger than one, there can be mixed Coulomb-Higgs branches where part of the gauge group is broken to a diagonal torus (along those components BPS solutions describe a deformed Coulomb branch) and part is completely broken (admitting vortex solutions).

\subsection{Computation of the partition function}

Given the various classes of solutions to the BPS equations found in the previous section, the computation of the partition function requires two more steps: the evaluation of the classical action and of the one-loop determinant of quadratic fluctuations around the BPS configurations, and the sum/integration over the space of BPS configurations.

\subsubsection{One-loop determinants from an index theorem}
\label{sec: 1-loop dets}

For the computation of the one-loop determinants around non-constant configurations, one most conveniently makes use of an equivariant index theorem for transversally elliptic operators \cite{Atiyah:1974}, as in \cite{Drukker:2012sr}. A similar technique has been used on $S^4$ \cite{Pestun:2007rz, Gomis:2011pf} and $S^2$ \cite{Benini:2012ui}. One can give a cohomological form to the $\cQ$-exact localizing action (this point is well explained in \cite{Pestun:2007rz, Gomis:2011pf}), and, with the equivariant index theorem, the one-loop determinants of quadratic fluctuations only get contributions from the fixed points of the equivariant rotations on the worldvolume. Recall that the localizing supercharge squares to
\be
\label{qsqoneloop}
\mathcal{Q}^2 = \cL^A_\xi  -\fS - \frac{i}{2}\left(\frac{1}{\ell} + \frac{1}{\tilde{\ell}}\right) R \;.
\ee
The vector field $\xi = \frac1\ell \partial_\varphi + \frac1{\tilde\ell} \partial_\chi$ does not have fixed points on $S^3_b$, on the other hand its orbits do not close for generic values of $b$ ($\xi$ generates a non-compact isometry group $\bR$) and since the index theorem requires a compact group action, we cannot use it directly.%
\footnote{For special values of $b$, \eg{} the round sphere $b=1$, the group action is a compact $U(1)$. Still the index theorem determines the index up to torsion, and in fact in those cases the index turns out to be pure torsion. We thank Takuya Okuda for correspondence on this issue.}
The idea of \cite{Drukker:2012sr} is to write $\xi = \big( \frac1\ell + \frac1{\tilde\ell} \big) \partial_{\psi_H} + \big(\frac1\ell - \frac1{\tilde\ell} \big)\partial_{\phi_H}$ in Hopf coordinates: it generates a free rotation of the Hopf fiber and a rotation of the base space. We can reduce the operator for quadratic fluctuations (\ie{} the operator resulting from the quadratic expansion of the localizing action around the background) along the Hopf fiber, obtaining a transversally elliptic operator on the base $S^2$. We thus reduce the problem to the computation of a one-loop determinant on the base $S^2$, dressed by the KK modes on the Hopf fiber. The projection of $\xi$ to $S^2$ gives a rotation with fixed points at $\theta = 0$ (which we call North) and $\theta = \frac{\pi}{2}$ (which we call South). This is exactly the setup in \cite{Benini:2012ui}. Identifying the equivariant parameters of the $U(1)_{\partial_{\phi_H}}\times U(1)_R \times G$ action as $\varepsilon = \frac1\ell - \frac1{\tilde\ell}$,  $\check\varepsilon = \frac{1}{\ell} + \frac{1}{\tilde{\ell}}$ and $a = -i\big( \frac{1}{\ell} A_\varphi + \frac{1}{\tilde{\ell}} A_\chi \big) -\fS$, following \cite{Drukker:2012sr} we obtain (see appendix \ref{sec: one-loop det}) the one-loop determinant for a chiral multiplet of R-charge $q$ in gauge representation $\cR$:
\be
\label{1loopchiral long}
Z_\text{1-loop}^\text{chiral} \text{``}=\text{''} \prod_{w \in \cR} \prod_{n \in \bZ} \prod_{m \geq 0} \, \frac{(m+1) \ell^{-1} + n\tilde\ell^{-1} - \frac q2 \check\varepsilon - i\, w(a_S)}{n \ell^{-1} - m \tilde\ell^{-1} - \frac q2 \check\varepsilon - i\, w(a_N)} \;.
\ee
In all BPS configurations that we consider in this section, $a_N = a_S \equiv a$ and some further simplifications take place. It is also convenient to introduce the rescaled variable $\hat a \equiv \sqrt{\ell\tilde\ell}\, a$, as well as $b \equiv \sqrt{\tilde\ell/\ell}$ and $Q = b + b^{-1}$. Rescaling numerator and denominator of (\ref{1loopchiral long}) by $\sqrt{\ell\tilde\ell}$ and neglecting overall signs, we are led to
\be
\label{1loopchiral}
Z_\text{1-loop}^\text{chiral} \text{``}=\text{''} \prod_{w \in \cR} \prod_{m,n\geq 0} \, \frac{mb + nb^{-1} + \big(1-\frac q2)Q - i w(\hat a)}{mb + nb^{-1} + \frac q2 Q + iw(\hat a)} = \prod_{w\in\cR} s_b \bigg( \frac{iQ}2 (1-q) + w(\hat a) \bigg) \;.
\ee
The last one is the regulated expression found in \cite{Hama:2011ea}, in terms of the double sine function $s_b$. The one-loop determinant for the vector multiplet is simply
\be
\label{1loopvec}
Z_\text{1-loop}^\text{vec} = \prod_{\alpha >0} 2\sinh \big(\pi b \, \alpha(\hat a) \big) \, 2\sinh\big( \pi b^{-1} \, \alpha(\hat a) \big) \;,
\ee
where the product is over the positive roots $\alpha$ of the gauge group.

\subsubsection{Coulomb branch}
\label{sec: S3 Coulomb branch}

Let us first quickly review the Coulomb branch localization formula, obtained by choosing $\zeta = 0$ in $H(\phi)$, or taking positive R-charges. The matrix model was derived in \cite{Hama:2011ea}. The only $\cQ$-closed but not $\cQ$-exact pieces of classical action are the CS and FI terms (that we report in appendix \ref{app: SUSY actions}). Evaluation on the Coulomb branch configurations gives
\be
S_\text{cl} =  i \pi \Tr_{CS} \hat\sigma^2 - 2\pi i \Tr_{FI} \hat\sigma \; ,
\ee
in terms of the rescaled adjoint scalar $\hat\sigma \equiv \sqrt{\ell \tilde\ell} \, \sigma$. The weighted traces $\Tr_{CS}$ and $\Tr_{FI}$ are spelled out in appendix \ref{app: SUSY actions}, and for $U(N)$ at level $k$ they reduce to $S_\text{cl} =  i \pi k \Tr \hat\sigma^2 - 2\pi i \xi \Tr \hat\sigma$.

Since the equivariant parameters for gauge transformations are equal at the two fixed circles, $\hat{a}_N = \hat{a}_S = -\hat\sigma$, the one-loop determinants (\ref{1loopchiral}) and (\ref{1loopvec}) are
\be
\label{Coulomb 1-loops}
Z_\text{1-loop}^\text{chiral} = \prod_{w\in\cR} s_b\bigg( \frac{iQ}2 (1-q) - w(\hat\sigma) \bigg) \;,\qquad\qquad
Z_\text{1-loop}^\text{vec}  = \prod_{\alpha>0} 2\sinh \big( \pi b \, \alpha(\hat\sigma) \big) \, 2\sinh \big( \pi b^{-1} \, \alpha(\hat\sigma) \big) \;.
\ee
This leads to the matrix integral of \cite{Hama:2011ea}:
\be
\label{Coulomb partition function}
Z_{S^3_b} = \frac1{|\cW|} \int \bigg( \prod_{a=1}^{\rank G} d\hat\sigma_a \bigg) \; e^{- i \pi \Tr_{CS} \hat\sigma^2 + 2\pi i \Tr_{FI} \hat\sigma} \; Z_\text{1-loop}^\text{vec} \; Z_\text{1-loop}^\text{chiral} \;,
\ee
where $|\cW|$ is the dimension of the Weyl group. Notice that the Vandermonde determinant for integration over the gauge algebra $\fg$ cancels against the one-loop determinant for gauge-fixing ghosts.

\subsubsection{Deformed Coulomb branch}
\label{sec: def Coulomb branch S3}

Let us now study the contributions for $\zeta \neq 0$.
The classical CS and FI actions evaluated on the deformed Coulomb branch configurations give
\be
\label{deformed actions}
S_\text{cl}^{CS} = i\pi \Tr_{CS} \big( \hat\sigma - i \zeta \kappa \big)^2 \;,\qquad\qquad
S_\text{cl}^{FI} = -2\pi i \Tr_{FI} \big( \hat \sigma - i \zeta \kappa \big) \;,
\ee
and we defined the constant
\be
\kappa \,\equiv\, \frac{\vol(S^3_b)}{4\pi^2r} = \frac{r^2}3 \big(Q - Q^{-1} \big) = \frac r3 \, \frac{\ell^2 + \ell\tilde\ell + \tilde\ell^2}{\ell + \tilde\ell} \;,
\ee
where $r\equiv\sqrt{\ell\tilde\ell}.$ In both cases the effect of the deformation parameter $\zeta$ is effectively to shift the integration variable $\hat\sigma$ in the imaginary direction. The same shift occurs in the equivariant gauge parameters
$$
\hat a_N = \hat a_S = - \hat\fS + i\zeta \kappa
$$
defined above (\ref{1loopchiral long}), as it follows from (\ref{deformedgaugefield}), and so also the one-loop determinants simply suffer an effective imaginary shift of $\hat\sigma$. Therefore the whole deformed Coulomb branch contribution is simply obtained from the undeformed Coulomb branch expression (\ref{Coulomb partition function}) by shifting the integration contours in the imaginary directions.

Since the parameter $\zeta$ was introduced via a $\cQ$-exact term in the action, the partition function should not depend on it. For $\zeta = 0$ we have the original Coulomb branch integral (\ref{Coulomb partition function}). Upon turning on $\zeta$ we effectively deform the contours, shifting them in the imaginary directions, and the integral remains constant until we cross some pole of the chiral one-loop determinant. One can anticipate what happens when crossing a pole based on the bound \eqref{bound}: the imaginary coordinates of the poles precisely correspond to values of $\zeta$ for which new vortices appear on $S^3_b$ as solutions to the vortex equations, and the contribution from the vortices precisely accounts for the jumps in the deformed Coulomb branch integral.

\paragraph{Suppression.}
Our goal is to derive a localization procedure that reduces the partition function to a pure sum over vortices, with no spurious contributions from deformed Coulomb branches. In order to do that, we can take a suitable limit $\zeta_a \to \pm \infty$: in favorable situations, there exists (for a choice of signs) a limit in which the deformed Coulomb branch contribution vanishes.

Let us define the $U(1)$ charges of a gauge representation $\cR_j$: $\fq^{(a)}_j \equiv w(h_a)$, where $h_a$ are the Cartan generators of the Abelian factors in the gauge group, as in (\ref{expansion fake FI}), while $w$ is any one weight of $\cR_j$.%
\footnote{There is no dependence on the particular weight $w$ chosen, since the $U(1)$ generators commute with all roots of the simple factors.}
We also decompose $\hat\sigma = \hat\sigma_R - i\zeta\kappa$ into its real and imaginary parts.
Using the asymptotic behavior of the double-sine function (see \eg{} the appendix of \cite{Bytsko:2006ut}):
\bea
s_b(z) \;\to\; \left\{ \begin{aligned} &e^{+ i \frac\pi2 \big( z^2 + \frac1{12}(b^2 + b^{-2}) \big)} \qquad && |z| \to \infty \;,\quad |\arg z| < \frac\pi 2 \\
& e^{-i\frac\pi2 \big( z^2 + \frac1{12}(b^2 + b^{-2}) \big)} \qquad && |z| \to \infty \;,\quad |\arg z| > \frac\pi 2 \;, \end{aligned} \right.
\eea
one finds that the absolute value of the integrand in the partition function matrix model has the following suppression factor, for $\zeta_a \to \pm \infty$:
$$
\Big| \text{integrand} \Big| \;\sim\; \exp \Bigg[ -2\pi \kappa \sum_a \zeta_a \bigg( \Tr_{CS}\hat\sigma_R h_a - \Tr_{FI} h_a + \frac12 \sum_{\cR_j} \fq_j^{(a)} \sum_{w\in \cR_j} \big| w(\hat\sigma_R) + m_j \big| \bigg) \Bigg] \;,
$$
where the first two terms in parenthesis originate from the classical action while the last term comes from the chiral multiplets in those representations $\cR_j$ with $\fq_j^{(a)} \neq 0$. The one-loop determinants of chiral multiplets with $\fq_j^{(a)}=0$ and that of vector multiplets are unaffected by $\zeta$. One can achieve a suppression of the deformed Coulomb branch contribution if there exists a choice of signs in the limit $\zeta_a \to \pm\infty$ such that the factor above goes to zero for all values of all components of $\hat\sigma_R$.

As a concrete example, consider a $U(N)$ theory with $N_f$ fundamentals, $N_a$ antifundamentals and some adjoint chiral multiplets (there is a single Abelian factor in the gauge group, and $\fq$ equals $1$, $-1$ and $0$ respectively). Setting the real masses to zero for simplicity, the factor above provides a suppression of the deformed Coulomb branch for
\be
\zeta \to +\infty \qquad\qquad\text{and}\qquad\qquad
- \frac{N_f - N_a}2 < k < \frac{N_f - N_a}2 \;,
\ee
in particular $N_f > N_a$, where the two constraints come from positive and negative $\hat\sigma_R$. Similarly, we have suppression for
\be
\zeta \to -\infty \qquad\qquad\text{and}\qquad\qquad
\frac{N_f - N_a}2 < k < - \frac{N_f - N_a}2 \;,
\ee
In particular $N_a > N_f$. These two cases, $|k| < |N_f - N_a|/2$, are the ``maximally chiral'' theories of \cite{Benini:2011mf}. In case one or both bounds are saturated, then the true FI term $\xi$ needs to have the correct sign.

We stress that if the ``maximally chiral'' condition (including saturations of the bounds) is not met, \ie{} if $|k| \leq |N_f-N_a|/2$ is not met, the deformed Coulomb branch contribution is \emph{not} suppressed. As we will see in the next section, this translates to the fact that the Coulomb branch integral cannot be closed neither in the upper nor lower half-plane, and reduction to a sum over residues (as in \cite{Pasquetti:2011fj}) requires some more clever procedure (if possible at all).

\subsubsection{Higgs branch and vortex partition function}
\label{sec: VPF S3}

For finite values of the deformation parameters $\zeta_a$, among the BPS configurations of section \ref{sec: solutionsS3} we find Higgs vacua and vortex solutions, where the (anti)vortex numbers $(m,n)$ are bounded by (\ref{bound}) (or its multi-dimensional version). These BPS configurations contribute to the path integral, besides the deformed Coulomb branch discussed before. Let us determine their contribution.

The classical actions can be integrated exactly (even though the vortex solutions cannot be written explicitly) using $D = - \sigma/f + iH(\phi)$, the BPS equations (\ref{BPS eqns components YM}) and the knowledge of $A_\varphi(\theta)$ at $\theta = 0,\frac\pi2$ in a globally defined gauge with $A_\theta = 0$, as discussed around (\ref{close to core}). One finds
\be
\label{classical actions vortices}
S_\text{cl}^{CS} = i\pi \Tr_{CS} \big( \hat\sigma - i b^{-1}m - i bn \big)^2 \;,\qquad\qquad
S_\text{cl}^{FI} = -2\pi i \Tr_{FI} \big( \hat\sigma - i b^{-1}m - i bn \big) \;.
\ee
Here the vortex numbers $m,n$ should really be thought of as GNO quantized \cite{Goddard:1976qe} elements of the gauge algebra, \ie{} belonging to the coweight lattice.

The evaluation of the one-loop determinants for the off-diagonal W-bosons and all chiral multiplets not getting a VEV is straightforward: one identifies the equivariant gauge transformation parameters in the vortex background from the expression of $\cQ^2$ at the poles:
\be
\hat{a}_N = \hat{a}_S = - \big( \hat\fS - i b^{-1} m - i b n \big) \;.
\ee
These values have to be plugged into \eqref{1loopchiral} and (\ref{1loopvec}). For the $\rank G$ chiral multiplets that get a VEV and, by Higgs mechanism, pair with the vector multiplets along the maximal torus of the gauge group becoming massive, one has to be more careful. As pointed out in \cite{Doroud:2012xw}, the one-loop determinant for the combined system is just the residue of the chiral one-loop.%
\footnote{The chiral one-loop diverges because it is evaluated at a point on the Coulomb branch where the chiral multiplet, before pairing with the vector multiplet, is massless. Taking the residue corresponds to removing the zero-mode.}
Therefore the total contribution from the chiral multiplets is
\be
\label{residue prescription}
Z_\text{1-loop}^\text{chiral} = \Res_{\fS \,\to\, \fS_{H}} \prod_{w \,\in\, \fR} s_b \bigg( \frac{iQ}{2} - w \big(\hat\fS - i m b^{-1} - inb \big) \bigg) \;,
\ee
where $\fS_{H}$ denotes $\fS$ evaluated on the particular Higgs vacuum, and the R-charges have been set to zero. Finally, since each BPS solution is a smooth configuration with no moduli, we simply sum over them with weight 1.

From (\ref{residue prescription}) it is clear that the sum of the contributions from the finite number of vortices satisfying the bound (\ref{bound}) exactly accounts for the jumps in the deformed Coulomb branch contribution every time the integration contour---which is shifted in the imaginary directions by $\zeta_a$---crosses a pole of the chiral one-loop determinant. This of course is expected, since the path integral should not depend on $\zeta$.

\paragraph{Vortex partition function.}
We obtain a more interesting result if we take a suitable $\zeta_a \to \pm\infty$ limit in which the deformed Coulomb branch contribution vanishes, and there is no bound on the (anti)vortex numbers. Conditions for the existence of such a limit were discussed in section \ref{sec: def Coulomb branch S3}.

In this limit the path integral is completely dominated by (anti)vortex-string configurations wrapping the northern and southern circles, and whose size shrinks to zero. The resummed contribution of all vortex strings is accounted by the K-theoretic vortex partition function, $Z_\text{vortex}$, which can be computed on the twisted $\bR^2_\epsilon \times S^1$: $\bR^2$ is rotated by the equivariant parameter $\epsilon$ as we go around $S^1$, and this effectively compactifies the space. In fact one associates equivariant parameters to flavor symmetries as well. In a suitable scaling limit in which $S^1$ shrinks (together with the equivariant parameters), one recovers the vortex partition function in $\Omega$-background of \cite{Shadchin:2006yz}. This all is the 2d analog of the 4d and 5d instanton partition functions constructed in \cite{Nekrasov:2002qd, Nekrasov:2003rj}.

Let us compute the partition function in this limit. First, we have a finite number of Higgs vacua. In each vacuum, $\hat\sigma_\alpha$ are fixed to some specific (real) values that are functions of the real masses. The classical actions (\ref{classical actions vortices}) provide a weighting factor to $Z_\text{vortex}$ for the vortex configurations, times an overall classical contribution:
\be
S_\text{cl} = i\pi \Tr_{CS} \hat\sigma^2 - 2\pi i \Tr_{FI} \hat\sigma \;.
\ee
The weighting factors for (anti)vortices have a term quadratic in the vortex number and a linear term:
\bea
e^{-S_\text{v}} &= \exp\big[ i\pi b^{-2} \Tr_{CS} m^2 + 2\pi b^{-1}\big( - \Tr_{CS} \hat\sigma \cdot\, + \Tr_{FI} \big) m \big] \\
e^{-S_\text{av}} &= \exp\big[ i\pi b^2 \Tr_{CS} n^2 + 2\pi b\big( - \Tr_{CS} \hat\sigma \cdot\, + \Tr_{FI} \big) n \big] \;.
\eea
The actions (\ref{classical actions vortices}) also give rise to a term $e^{2\pi i \Tr_{CS} m n}$: in the absence of parity anomaly in the matter sector, $\Tr_{CS} mn$ is integer and the term equals 1; otherwise $\Tr_{CS}$ is semi-integer and such that the term is a sign precisely canceling the parity anomaly.%
\footnote{Concretely, for $U(N)_k$ with $N_f$ fundamentals and $N_a$ antifundamentals, cancelation of the parity anomaly requires $2k + N_f - N_a \in \bZ$. The general case is discussed in \cite{Aharony:1997bx}.}

Second, the one-loop determinants for the vector multiplet and the chiral multiplets not acquiring a VEV are as in (\ref{Coulomb 1-loops}). The $\rank G$ chiral multiplets acquiring VEV bring a residue factor, which in this case is just 1. Finally, the vortex partition function $Z_\text{vortex}$ depends on equivariant parameters for rotations of $\bR^2$ ($\varepsilon$) and flavor rotations ($g$): they are identified---at $\theta=0$ (N) and $\theta = \frac\pi2$ (S)---from the $SU(1|1)$ complex of the supercharge $\cQ$ at the poles, \ie{} from $\cQ^2$ in (\ref{qsqoneloop}). We find
\be
\varepsilon_N = \frac{2\pi}{b^2} \;,\qquad g_N = -\frac{2\pi}{b}\hat\fS \;,\qquad\qquad\qquad
\varepsilon_S = 2\pi b^2 \;,\qquad g_S = -2\pi b\hat\fS \;.
\ee

Eventually, Higgs branch localization gives the following expression of the sphere partition function:
\be
\label{Higgs branch localization}
Z_{S^3_b} = \sum_{\text{Higgs vacua}} \; e^{-i\pi \Tr_{CS} \hat\sigma^2 + 2\pi i \Tr_{FI} \hat\sigma} \; Z'_\text{1-loop} \; Z_\text{v} \; Z_\text{av} \;.
\ee
The sum is over solutions to (\ref{Higgs branch equations}). The one-loop determinant $Z'_\text{1-loop}$ does not contain the $\rank G$ chiral multiplets getting VEV in (\ref{Higgs branch equations}). The (anti)vortex-string contributions are expressed in terms of the 3d vortex partition function:
\bea
\label{parameter map S3}
Z_\text{v} &= Z_\text{vortex} \Big( e^{i\pi b^{-2} \Tr_{CS} \cdot} \,,\, e^{2\pi b^{-1} (-\Tr_{CS} \hat\sigma\cdot\, + \Tr_{FI} \cdot)} \,,\, \frac{2\pi}{b^2} \,,\, - \frac{2\pi}b \hat\fS \Big) \\
Z_\text{av} &= Z_\text{vortex} \Big( e^{i\pi b^2 \Tr_{CS} \cdot} \,,\, e^{2\pi b (-\Tr_{CS} \hat\sigma\cdot\, + \Tr_{FI} \cdot)} \,,\, 2\pi b^2 \,,\, - 2\pi b \hat\fS \Big) \;.
\eea
The first two arguments in the vortex partition function are exponentiated linear functions on the gauge algebra, corresponding to the quadratic and linear weights for the vortex numbers; the third is the rotational equivariant parameter and the last one includes all flavor equivariant parameters. Notice that the expression (\ref{Higgs branch localization}) is very much in the spirit of the ``holomorphic blocks'' of \cite{Beem:2012mb}.

We shall give a concrete example in the next section.

\subsection{Matching with the Coulomb branch integral}
\label{sec: Rewriting Coulomb branch}

We would like to briefly show, in the simple example of a $U(N)$ gauge theory with $N_f$ fundamentals and $N_a$ antifundamentals, that Higgs branch and Coulomb branch localization produce in fact the same partition function, written in a completely different way. This computation has already been done in the case of $U(1)$ in \cite{Pasquetti:2011fj}, and in the case of $U(N)$ in \cite{Taki:2013opa}, therefore we will just review it in our conventions. We stress, however, that this computation is only valid for
\be
|k|\leq \frac{|N_f - N_a|}2
\ee
where $k$ is the Chern-Simons level; these theories have been dubbed ``maximally chiral'' in \cite{Benini:2011mf}.%
\footnote{\label{foo: max/min chiral}
The computation in this section leads to an expression for the $S^3$ partition function which identically vanish for $\max(N_f,N_a) < N$, signaling supersymmetry breaking. The fact that the maximally chiral theories ($|k| \leq \frac{|N_f - N_a|}2$) break supersymmetry for $\max(N_f,N_a)<N$ has been noticed in \cite{Benini:2011mf}. On the other hand, the minimally chiral theories ($|k| \geq \frac{|N_f - N_a|}2$) generically do \emph{not} break supersymmetry for $\max(N_f,N_a)<N$; the simplest example is pure Chern-Simons theory. In fact the manipulations carried out here are not valid in the latter case. A similar reasoning applies to the index $Z_{S^2 \times S^1}$. We thank Ofer Aharony for this observation.}

The theory has $SU(N_f) \times SU(N_a) \times U(1)_A$ flavor symmetry. We will use a ``quiver'' notation, in which the fundamentals are in the antifundamental representation of the flavor group $SU(N_f)$, and viceversa. Then we can introduce real masses $m_\alpha$ for fundamentals and $\tilde m_\beta$ for antifundamentals, defined up to a common shift (which corresponds to a shift of the adjoint scalar $\sigma$). Generic positive R-charges are encoded as imaginary parts of the masses.

The matrix integral (\ref{Coulomb partition function}) is given by (we removed hat from $\hat\sigma$):
\begin{multline}
Z_{S^3_b}^{U(N), N_f,N_a} = \frac1{N!}\int d^N\sigma \, e^{-i\pi k \sum \sigma_i^2 + 2\pi i \xi \sum \sigma_i} \prod_{i < j}^N 4 \sinh \big(  \pi b^{-1} (\sigma_i - \sigma_j ) \big)  \sinh \big( \pi b  (\sigma_i - \sigma_j ) \big) \\
\times \, \prod_{i=1}^N \, \frac{\prod_{\beta=1}^{N_a} s_b \left( \frac{iQ}{2} + \sigma_i - \tilde m_\beta \right) }{\prod_{\alpha=1}^{N_f} s_b \left( -\frac{iQ}{2} + \sigma_i - m_\alpha \right)} \;,
\end{multline}
where we used $s_b(-x) = s_b^{-1}(x).$ Our goal is to rewrite it as a sum over residues, as done in \cite{Pasquetti:2011fj, Taki:2013opa}. First, one can employ twice the Cauchy determinant formula that we use in the following form:
\be
\label{Cauchy}
\prod_{i<j}^N 2\sinh(x_i-x_j) = \frac1{\displaystyle \prod\nolimits_{i<j}^N 2\sinh(\chi_i-\chi_j)} \quad \sum_{s \in S^N}(-1)^s \prod_{i=1}^N \prod_{j \neq s(i)}^N 2\cosh(x_i-\chi_j) \;,
\ee
where the auxiliary variables $\chi_i$ must satisfy $\chi_i \neq \chi_j \pmod{\pi i}$, to separate the interacting matrix-model into a product of simple integrals. The simple integrals will contain two sets of auxiliary variables $\chi_i$, $\tilde\chi_i$.
Assuming that  $|k| < \frac{N_f-N_a}2$ (or $|k| \leq \frac{N_f - N_a}2$ and $\xi<0$), these integrals can be  computed by closing the contour in the lower-half plane and then picking up the residues. The regime $|k| \leq \frac{N_a - N_f}2$ can be studied in a similar way, closing the contours in the upper-half plane. One gets contributions from the simple poles of the one-loop determinants of fundamentals, located at the zeros of $s_b$ in the denominator:  $\sigma_j = m_{\gamma_j}  - i \mu_j b - i \nu_j b^{-1} \,\equiv\, \tau_j(m_{\gamma_j},\mu_j,\nu_j)$ for $\mu_j,\nu_j \in \bZ_{\geq 0}$ and ${\gamma_j}  = 1,\ldots,N_f$.
Applying the Cauchy determinant formula backwards, to re-absorb the auxiliary variables, one obtains
\begin{multline}
\label{Zrewritten}
Z_{S^3_b} = \frac{(-2\pi i)^N}{N!} \; \sum_{\vec\gamma \,\in\, (\bZ_{N_f})^N} \; \sum_{\vec\mu,\, \vec \nu \,\in\, \bZ_{\geq0}^N} \; e^{-i\pi k \sum \tau_i^2 + 2\pi i \xi \sum \tau_i} \;
\prod_{i<j}^N 4\sinh \big( \pi b(\tau_i -\tau_j ) \big) \sinh \big( \pi b^{-1}(\tau_i - \tau_j ) \big) \\
\times \prod_{i=1}^N \Bigg( \frac{\prod_{\beta=1}^{N_a} s_b \left( \frac{iQ}2 + \tau_i - \tilde m_\beta \right) }{\prod_{\alpha\neq \gamma_i}^{N_f} s_b \left( -\frac{iQ}2 + \tau_i  - m_\alpha \right)} \; \Res_{x\to 0} s_b \Big( \frac{iQ}2 + i\mu_ib + i\nu_ib^{-1} -x \Big) \Bigg) \;.
\end{multline}
Of course, one could have just collected the residues of the multi-dimensional integral with no need of the Cauchy formula. The residue can be computed with the identity
\be
\label{identity}
s_b \Big( x + \frac{iQ}2 + i\mu b + i\nu b^{-1} \Big) = \frac{(-1)^{\mu\nu} \; s_b \Big( x + \dfrac{iQ}2 \Big)}{\prod_{\lambda=1}^{\mu} 2 i \sinh \pi b \big( x + i\lambda b \big) \phantom{\Big|}\; \prod_{\kappa = 1}^{\nu} 2 i \sinh \pi b^{-1} \big(x + i\kappa b^{-1} \big) }
\ee
and $\Res_{x\to0} s_b\big( x + iQ/2\big) = 1/2\pi i$. At this point one can factorize the summation into a factor independent of $\vec\mu$ and $\vec\nu$, a summation over $\vec\mu$ and a summation over $\vec\nu$. To achieve that one uses $2k + N_f-N_a = 0 \pmod{2}$, which is the condition for parity anomaly cancelation, so that $(-1)^{(N_f-N_a+2 k)\sum_i\mu_i\nu_i}=1$. Finally one observes that each of the two summations over $\vec\mu$ and $\vec\nu$ vanishes if we choose $\gamma_i = \gamma_j$ for some $i,j$, and on the other hand it is symmetric under permutations of the $\gamma_i$'s. Therefore we can restrict the sum over unordered combinations $\vec\gamma \in C(N,N_f)$ of $N$ out of the $N_f$ flavors, and cancel the $N!$ in the denominator.

We can also use the following identity (see \eg{} appendix B of \cite{Hwang:2012jh}), valid when the $\gamma_i$'s are distinct:
\begin{align}
\label{identitysinh}
& \frac{\prod_{j<k}^N \sinh\left(X_{\gamma_k} - X_{\gamma_j} + i (\mu_k - \mu_j) Y \right) }{\prod_{i= 1}^{N} \prod_{\beta=1}^{N_f}\prod_{\lambda = 1}^{\mu_i} \sinh\left(  X_{\gamma_i} - X_{\beta} +i \lambda Y \right) } = \\
&\qquad = \frac{(-1)^{\sum_j \mu_j} \prod_{j<k}^N \sinh\left(X_{\gamma_k} - X_{\gamma_j} \right) }{\prod_{k=1}^{N} \prod_{\lambda=1}^{\mu_k} \left[ \prod_{j=1}^{N}  \sinh\left(X_{\gamma_k}-X_{\gamma_j} - i (\mu_j - \lambda+1)Y \right) \right] \left[ \prod_{\beta \not\in \{\gamma_l\} }^{N_f} \sinh\left( X_{\gamma_i} - X_\beta + i \lambda Y \right)\right] } \nn
\end{align}
and the observation $\prod_{i<j}^N (-1)^{\mu_i - \mu_j} = (-1)^{(N-1)\sum_i \mu_i}$, to eventually write:
\be
Z_{S^3_b} = \sum_{\vec\gamma \,\in\, C(N,N_f)} Z_\text{cl}^{(\vec\gamma)} \; Z_\text{1-loop}^{\prime\,(\vec\gamma)} \; Z_\text{v}^{(\vec\gamma)} \; Z_\text{av}^{(\vec\gamma)} \;,
\ee
which exactly matches with the general result of Higgs branch localization (\ref{Higgs branch localization}). The summation is over classical Higgs vacua, \ie{} over solutions to the algebraic D-term equations (\ref{Higgs branch equations}). Then we have a simple classical piece, the one-loop determinant of all fields except the $N$ chiral multiplets (specified by $\vec\gamma$) getting a VEV and Higgsing the gauge group, the vortex and the anti-vortex contributions; all these functions are evaluated at the point $(\vec\gamma)$ on the Coulomb branch solving the D-term equations. Using a notation in which $\alpha\in\vec\gamma$ denotes the flavor indices in the combination $\vec\gamma$, we can write the classical and one-loop contributions as
\bea
Z_\text{cl}^{(\vec\gamma)} &= \prod\nolimits_{\alpha\in\vec\gamma} \; e^{-i\pi k m_\alpha^2 + 2\pi i \xi m_\alpha} \\
Z_\text{1-loop}^{\prime\,(\vec\gamma)} &= \prod_{i \in \vec\gamma} \frac{ \prod_{\beta=1}^{N_a} s_b \left( \frac{iQ}2 + m_i - \tilde m_\beta \right)}{ \prod_{\alpha\, (\neq i)}^{N_f} s_b \left( - \frac{iQ}2 + m_i - m_\alpha \right)} \cdot \prod_{\substack{i,j \in \vec\gamma \\ i\neq j}} 4 \sinh \big( \pi b (m_i - m_j) \big) \sinh \big( \pi b^{-1} (m_i - m_j) \big) \;,
\eea
the (anti)vortex contributions as
\bea
Z_\text{v}^{(\vec\gamma)} &= Z_\text{vortex}^{(\vec\gamma)}\Big( e^{i\pi b^{-2}k} \,,\, e^{2\pi b^{-1}(-km_j + \xi)} \big|_{j\in\vec\gamma} \,,\, \frac{2\pi}{b^2} \,,\, - \frac{2\pi}b m_\alpha \,,\, - \frac{2\pi}b \tilde m_\beta \Big) \\[.4em]
Z_\text{av}^{(\vec\gamma)} &= Z_\text{vortex}^{(\vec\gamma)}\Big( e^{i\pi b^2 k} \,,\, e^{2\pi b(-km_j + \xi)} \big|_{j\in\vec\gamma} \,,\, 2\pi b^2 \,,\, - 2\pi b m_\alpha \,,\, - 2\pi b \tilde m_\beta \Big) \;,
\eea
and the vortex-string partition function turns out to be (for $N_f \geq N_a$):
\begin{multline}
\label{3d VPF}
Z_\text{vortex}^{(\vec\gamma)}\Big( Q_j \,,\, L_j \,,\, \varepsilon \,,\, a_\alpha \,,\, b_\beta \Big) = \sum_{\vec\mu \,\in\, \bZ_{\geq0}^N} \; \prod_{j \,\in\, \vec\gamma} \; Q_j^{\mu_j^2} \, L_j^{\mu_j} \; (-1)^{(N_f-N_a)\mu_j} \\
\times \prod_{\lambda=0}^{\mu_j-1} \frac{\displaystyle \prod\nolimits_{\beta=1}^{N_a} 2i\sinh \frac{a_j - b_\beta + i\varepsilon\lambda}2}{\displaystyle \prod_{l\,\in\,\vec\gamma} 2i\sinh \frac{a_j - a_l + i\varepsilon(\lambda - \mu_l)}2 \; \prod\nolimits_{\alpha\,\not\in\,\vec\gamma}^{N_f} 2i\sinh \frac{a_\alpha - a_j + i\varepsilon(\lambda - \mu_j)}2} \;.
\end{multline}
The map of parameters in $Z_\text{v}$ and $Z_\text{av}$ precisely agrees with our general expression (\ref{parameter map S3}). As we will see in section \ref{sec: matching S2S1}, precisely the same function $Z_\text{vortex}$ controls the partition function on $S^2 \times S^1$. Such an expression for $Z_\text{vortex}$ can be compared with \cite{Kim:2012uz}.%
\footnote{See also \cite{Chen:2013pha}.}

\subsection{Comparison with the two-dimensional vortex partition function}

Let us check that by taking the limit of small equivariant parameter and scaling at the same time all other parameters in the same way, the 3d vortex partition function (\ref{3d VPF}) reduces to the 2d vortex partition function. After a redefinition $\varepsilon \to - \varepsilon$, we take a limit $\varepsilon \to 0$ in (\ref{3d VPF}) keeping the ratios $a_\alpha/\varepsilon$ and $b_\beta/\varepsilon$ finite; we also send the CS level $k \to 0$, that corresponds to $Q_j \to 1$ and $L_j \to z$. We get
\be
Z_\text{vortex}^{(\vec\gamma)} \,\to\, \sum_{\vec\mu \,\in\, \bZ_{\geq0}^N} \frac{z^{|\vec\mu|}}{(-\varepsilon)^{(N_f - N_a)|\vec\mu|}} \; \prod_{j\,\in\,\vec\gamma} \; \frac{\displaystyle \prod\nolimits_{\beta=1}^{N_a} \left( \frac{ia_j - ib_\beta}\varepsilon \right)_{\mu_j} }{ \displaystyle \prod_{l\,\in\,\vec\gamma} \left( \frac{ia_j - ia_l}\varepsilon - \mu_l \right)_{\mu_j} \; \prod\nolimits_{\alpha\,\not\in\,\vec\gamma}^{N_f} \left( \frac{ia_\alpha - ia_j}\varepsilon - \mu_j \right)_{\mu_j} } \;.
\ee
Here $|\vec\mu| = \sum_j \mu_j$ and we used the Pochhammer symbol $(a)_n = \prod_{k=0}^{n-1}(a+k)$. This expression is precisely the standard two-dimensional vortex partition function in $\Omega$-background, see \eg{} \cite{Benini:2012ui}.

\section{Higgs branch localization on $S^2\times S^1$}
\label{sec: S2xS1}

We will now move to the similar study of Higgs branch localization for $\cN=2$ theories on $S^2\times S^1$, whose path integral computes the three-dimensional supersymmetric index \cite{Bhattacharya:2008zy}. Localization on the Coulomb branch for $\cN=6$ Chern-Simons-matter theories was first performed in \cite{Kim:2009wb}, and later generalized to $\mathcal{N}=2$ theories in \cite{Imamura:2011su} (see also \cite{Kapustin:2011jm} for a further generalization in which magnetic fluxes for global symmetries are introduced). It was later pointed out in \cite{Dimofte:2011ju} (see also \cite{Aharony:2013dha}) that in the presence of non-trivial magnetic fluxes, the angular momentum of fields can be shifted by half-integer amounts, thus correcting the naive fermion number: such a different weighing of the magnetic sectors helps to verify various expected dualities.

The expression that results from Coulomb branch localization is a matrix integral over the holonomy of the gauge field. As in the previous section, we will perform an alternative Higgs branch localization, in which the relevant BPS configurations are discrete Higgs branches accompanied by towers of vortex strings at the north and south poles of the two-sphere.

\subsection{Killing spinors on $S^2 \times S^1$, supersymmetric index and deformed background}
\label{subsection:Killing spinors on S2S1 }

Supersymmetric theories on three-manifolds, among which $S^2\times S^1$, have been studied in \cite{Klare:2012gn, Closset:2012ru} considering the rigid limit of supergravity. In this approach, the auxiliary fields of the supergravity multiplet are treated as arbitrary background fields and SUSY backgrounds are found by setting to zero the gravitino variations; in the presence of flavor symmetries, one similarly sets to zero the external gaugino variations.

Here we will take a different approach: we will first recall the Killing spinor solutions on $S^2\times \bR$, and then compactify $\bR$ to $S^1$ with some twisted boundary conditions: the supersymmetric index with respect to the supercharges described by the Killing spinors indeed imposes twisted boundary conditions. In a path integral computation, however, the twisted boundary conditions are most conveniently described by turning on background fields for the charges appearing in the index formula, which finally leads to the desired theory on a deformed background.

We take the metric
\be
\label{metricS2S1}
ds^2 = r^2 (d\theta^2 + \sin^2\theta\, d\varphi^2) + d\tau^2 \;,
\ee
with vielbein $e^1 = r\, d\theta$, $e^2 = r\sin\theta\, d\varphi$, $e^3 = d\tau$, and set the background $U(1)_R$ field $V_\mu$ to zero. The spin connection is $\omega^{12} = - \cos\theta\, d\varphi$. Consider the Killing spinor equation
\be
\label{Killing spinor equation S2xS1}
D_\mu \epsilon = \gamma_\mu \hat\varepsilon
\ee
where $D_\mu = \partial_\mu + \frac14 \omega_\mu^{ab} \gamma_{ab}$. Following \cite{Imamura:2011su} we consider the factorized ansatz $\epsilon_\pm = f(\tau) \, \epsilon_\pm^{S^2}(\theta,\varphi)$, where the 2d spinor satisfies $D_{\hat\mu} \epsilon_\pm^{S^2} = \pm \frac1{2r} \gamma_{\hat\mu} \gamma^3 \epsilon_\pm^{S^2}$ with $\hat\mu=\theta,\varphi$. Plugging in (\ref{Killing spinor equation S2xS1}) gives
\be
\epsilon_\pm = e^{\pm\tau/2r} \, \epsilon_\pm^{S^2}(\theta,\varphi) \;,\qquad\qquad D_\mu \epsilon_\pm = \pm \frac1{2r} \gamma_\mu \gamma^3 \epsilon_\pm \;.
\ee
Notice that the spinors are not periodic on $S^1$ and twisted boundary conditions will be needed. On the sphere $S^2$ there are four Killing spinors; then we can write the $S^2 \times \bR$ spinors in a compact form as
\be
\epsilon_\pm = e^{\pm\tau/2r} \exp \Big( \mp \frac{i\theta}2 \, \gamma_2 \Big) \exp\Big( \frac{i\varphi}2\, \gamma_3 \Big) \, \epsilon_0
\ee
where $\epsilon_0 = \smat{C_1 \\ C_2}$ is constant.

\paragraph{Killing spinors for supersymmetric index.}
We will choose the spinor $\epsilon$ to be ``positive'' and with $\epsilon_0 = \smat{1\\0}$ (so that $\gamma_3\epsilon_0 = \epsilon_0$) and $\bar\epsilon$ to be ``negative'' and with $\bar\epsilon_0 = \smat{0\\1}$ (so that $\gamma_3\bar\epsilon_0 = - \bar\epsilon_0$):
\be
\label{KS}
\epsilon = e^{\tau/2r} e^{i \frac\varphi2} \mat{ \cos\theta/2 \\ \sin\theta/2} \;,\qquad\qquad \bar\epsilon = e^{-\tau/2r} e^{-i\frac\varphi2} \mat{ \sin\theta/2 \\ \cos\theta/2} \;.
\ee
Another useful spinor is
\be
\tilde \epsilon = -\bar\epsilon^c = i \, e^{-\tau/2r} e^{i \frac\varphi2} \mat{\cos\theta/2 \\ -\sin\theta/2}
\ee
which is also a ``negative'' Killing spinor.
We choose them of opposite positivity so that bilinears be independent of $\tau$; this also guarantees that there are no dilations in the algebra ($\rho = 0$). With these choices, the Killing vector and the functions appearing in the algebra are
\be
v^a = \bar\epsilon \gamma^a \epsilon = - \tilde\epsilon^\dag \gamma^a \epsilon = (0, \sin\theta, i) \;,\qquad
\bar\epsilon \epsilon = - \tilde\epsilon^\dag \epsilon = i \cos\theta \;,\qquad
\alpha = \frac 1r \bar\epsilon \gamma^3 \epsilon = \frac ir \;.
\ee
We also have
\be
\xi = i (\bar\epsilon \gamma^\mu \epsilon)\partial_\mu = \tfrac ir \partial_\varphi -\partial_\tau \;.
\ee
On the other hand $\epsilon^\dag\epsilon = e^{\tau/r}$ and $\tilde\epsilon^\dag\tilde\epsilon = e^{-\tau/r}$, as required by the dimension $\Delta$ (see below). The quantum numbers of the spinors are:
$$
\begin{array}{cccc}
\hline\hline
\text{Spinor} & \Delta & j_3 & R \\
\hline
\epsilon & -1/2 & 1/2 & -1 \\
\bar\epsilon & 1/2 & -1/2 & 1 \\
\hline
\end{array}
$$
obtained by acting with the operators $\Delta$ and $j_3$ as defined below; the R-charge follows from the supersymmetry variations. We also have
\be
[\delta_\epsilon, \delta_{\bar\epsilon} ] = \frac1r \Big( \big(\underbrace{-r\mathcal{L}_{\partial_\tau}^A}_{= \Delta}\big) - \big( \underbrace{-i\mathcal{L}_{\partial_\varphi}^A + r \cos\theta \sigma }_{= j_3}\big) - R \Big) \;.
\ee

\paragraph{Supersymmetric index and deformed background.}
The spinors are preserved by the mutually commuting operators
$$
\Delta - j_3 - R \;,\qquad\qquad R + 2j_3 \;.
$$
The first one is the commutator $[\delta_\epsilon, \delta_{\bar\epsilon}]$. We will compute the index
\be
\label{index}
I(x,\zeta_i) = \Tr \, (-1)^{2 j_3} e^{-\beta(\Delta - j_3 - R)} \, e^{-\xi(R + 2j_3)} \, e^{i \sum_j \mathfrak{z}_j F_j} \qquad\text{with}\qquad x = e^{-\xi} \;,\qquad \zeta_j = e^{i \mathfrak{z}_j} \;.
\ee
Here $F_j$ are the Cartan generators of the flavor symmetries and the circumference of $S^1$ is $\beta r$.
To correctly describe the fermion number in the presence of magnetic fluxes, we have used $2j_3$ \cite{Dimofte:2011ju, Aharony:2013dha}. Notice that convergence of the trace requires $|x|<1$.
For each Cartan generator of the flavor symmetry, besides the chemical potential $\zeta_j$ one could also turn on a fixed background flux on $S^2$ \cite{Kapustin:2011jm}: the only example we will consider in this paper is a flux for the topological symmetry $U(1)_J$.

In the path integral formulation on $S^2 \times S^1$, the index is described by the twisted periodicity conditions
\be\label{twistedpc}
\Phi(\tau + \beta r) = e^{\beta(-j_3 - R)} \; e^{\xi(R + 2j_3)} \; e^{-i \sum_j\mathfrak{z}_j F_j} \; \Phi(\tau) \;.
\ee
These are also the boundary conditions satisfied by the spinors (with $F_j=0$). By the field redefinition
$\tilde\Phi \equiv  e^{-\frac{\tau}{\beta r}\left(\beta(-j_3 - R)+\xi(R + 2j_3)-i \sum_j\mathfrak{z}_j F_j\right)} \Phi$, one can make the fields periodic again; such a redefinition is in fact a gauge transformation, indeed one can alternatively turn on background flat connections on $S^1$:
\be
V_\mu = \Big(0,0, -\frac ir + \frac{i\xi}{\beta r} \Big) \;,\qquad\qquad\qquad \tilde V_\mu^{(j)} = \Big( 0,0, \frac{\fz_j}{\beta r} \Big)
\ee
for the R- and flavor symmetries respectively. The twist by the rotational symmetry imposes the identification $(\tau,\varphi) \sim (\tau + \beta r, \varphi - i (\beta-2\xi))$. Introducing coordinates $\hat\tau = \tau$ and $\hat\varphi = \varphi + \frac{i (\beta-2\xi)}{\beta r}\tau$, the identification becomes $(\hat\tau,\hat\varphi) \sim (\hat\tau + \beta r, \hat\varphi)$. In hatted coordinates the metric \eqref{metricS2S1} is
\be
\label{deformedmetric}
ds^2 =  r^2 d\theta^2 + r^2 \sin^2\theta \, \Big[ d\hat\varphi - \frac ir \Big( 1 - \frac{2\xi}{\beta} \Big) \,  d\hat\tau \Big]^2 + d\hat\tau^2 \;,
\ee
which is complex. This metric can also be rewritten as
\be
\label{deformedmetric2}
ds^2 = r^2 d\theta^2 + \frac{r^2 \sin^2\theta}{1 - \big( 1-\frac{2\xi}{\beta} \big)^2 \sin^2\theta} d\hat\varphi^2 + \Big( 1 - \big(1 - \tfrac{2\xi}\beta \big)^2 \sin^2\theta \Big) \bigg( d\hat\tau - \frac{i r \big( 1-\frac{2\xi}{\beta} \big) \sin^2\theta }{1 - \big( 1 - \frac{2\xi}{\beta} \big)^2 \sin^2\theta} d\hat\varphi \bigg)^2 ,
\ee
which is a circle-fibration over a squashed two-sphere.

The index is thus computed by the partition function on a deformed background. A vielbein for \eqref{deformedmetric} is $e^1 = r\, d\theta, e^2 = r\sin\theta \big( d\hat\varphi - \frac ir \big( 1 - \frac{2\xi}\beta \big) d\hat\tau \big)$, $e^3 = d\hat\tau$, and the frame vectors are $e_1 = \frac1r \partial_\theta$, $e_2 = \frac{1}{r\sin\theta}\partial_{\hat\varphi}$, $e_3 = \partial_{\hat\tau} + \frac ir \big( 1 - \frac{2\xi}\beta \big) \partial_{\hat\varphi}$. The non-vanishing component of the spin connection is $\omega^{12} = - \cos\theta \big( d\hat\varphi - \frac ir \big( 1 - \frac{2\xi}\beta \big) d\hat\tau \big)$. The Killing spinors corresponding to \eqref{KS} are
\be
\label{twKS}
\epsilon = e^{i \hat\varphi/2} \mat{ \cos\theta/2 \\ \sin\theta/2} \;,\qquad\qquad \bar\epsilon =  e^{-i \hat\varphi/2} \mat{ \sin\theta/2 \\ \cos\theta/2} \;.
\ee
They satisfy $D_\mu \varepsilon = \frac{1}{2r} \gamma_\mu \gamma^3 \varepsilon$ and $D_\mu \bar\varepsilon = -\frac{1}{2r} \gamma_\mu \gamma^3 \bar\varepsilon$, where $D_\mu = \partial_\mu + \frac{1}{4}\omega_\mu^{ab}\gamma_{ab} - i V_\mu - i \sum_j \tilde V^{(j)}_\mu$, and $\tilde\epsilon = -\bar\epsilon^c$. The Killing vector and the functions appearing in the algebra are
\be
v^a = \bar\epsilon \gamma^a \epsilon = \big( 0, \sin\theta, i \big) \;,\qquad
\bar\epsilon \epsilon = i \cos\theta \;,\qquad
\alpha = \frac{i \xi}{\beta r} \;,\qquad \xi^\mu = i \bar\epsilon \gamma^\mu \epsilon = \Big( 0, \frac{2i\xi}{\beta r}, -1 \Big) \;.
\ee
We thus find
\be
[\delta_\epsilon, \delta_{\bar\epsilon} ] = -\mathcal{L}^A_{\partial_{\hat\tau}} + \frac{2i\xi}{\beta r} \mathcal{L}^A_{\partial_{\hat\varphi}} - \cos\theta\, \sigma - \frac{\xi}{\beta r} R + i \sum\nolimits_j \frac{\fz_j}{\beta r} F_j \;.
\ee

From standard arguments, it is known that the index is independent of the parameter $\beta$. A significant simplification takes place by setting $\beta = 2\xi$, since the rotational symmetry charge disappears from the trace \eqref{index}, and the complex metric \eqref{deformedmetric} becomes the real metric on the product space $S^2\times S^1$. Henceforth, we make this choice for the immaterial parameter $\beta$ and we further omit the hats.

\subsection{The BPS equations}

We will now proceed to derive the BPS equations. We define the quantities
\be
Y_a = W_a + \delta_{a3} \, \frac\sigma r \;,
\ee
where $W_a$ was defined in (\ref{def Ws}). Using the explicit expressions for the Killing spinors \eqref{twKS}, the BPS equations from the gaugino variations \eqref{gaugemultiplet} can be written as
\bea
0 &= \big(Y_3 + iD \big) \cos \frac{\theta}{2} + \big( D_1 \sigma - i Y_2 \big) \sin \frac{\theta}{2} \;,\qquad & 0 &= D_3 \sigma \cos \frac{\theta}{2} + \big( Y_1 -i D_2 \sigma \big)\sin \frac{\theta}{2} \\
0 &= \big(-Y_3 + iD\big) \sin \frac{\theta}{2} + \big( D_1 \sigma + i Y_2 \big) \cos \frac{\theta}{2} \;,\qquad & 0 &= -D_3 \sigma \sin \frac{\theta}{2} + \big( Y_1 +i D_2 \sigma \big)\cos \frac{\theta}{2} \;.
\eea

The localization locus can also be obtained from the positive definite deformation action (\ref{YM def action}), where now the action of $\ddagger$ is defined to be
\bea
(\mathcal{Q}\lambda)^{\ddagger} &= \epsilon^{\dagger} \Big( -\frac{1}{2} \gamma^{\mu\nu}F_{\mu\nu} - D - i\gamma^\mu D_\mu\sigma - \frac{i}{r} \sigma \gamma^3 \Big) = \epsilon^{\dagger} \big( - i \gamma^{r} (Y_r + D_r \sigma) - D \big) \\
(\mathcal{Q}\lambda^{\dagger})^{\ddagger} &=  \Big( \frac{1}{2} \gamma^{\mu\nu}F_{\mu\nu} + D - i\gamma^\mu D_\mu\sigma + \frac{i}{r} \sigma \gamma^3 \Big) \tilde{\epsilon} = \big( i (Y_r - D_r\sigma)\gamma^r + D   \big) \tilde{\epsilon} \;.
\eea
One then obtains $\cL^\text{def}_\text{YM} = \frac12 \Tr \big[ (Y_\mu)^2 + (D_\mu\sigma)^2 + D^2 \big]$.
Imposing the reality conditions, the Coulomb branch localization locus immediately follows:
\be
\label{BPSeqnsYM}
Y_\mu = 0 \;,\qquad\qquad D_\mu \sigma = 0 \;,\qquad\qquad D=0 \;.
\ee
Note that the string-like vortices are excluded by these equations since they imply $D_\mu F_{12}=0$.

Higgs branch localization can be achieved by adding another $\mathcal{Q}$-exact term to the deformation action. We use the same term as in (\ref{def action H}):
$$
\cL^\text{def}_\text{H} = \cQ \Tr \Big[ \frac{i (\epsilon^\dag \lambda - \lambda^\dag \tilde\epsilon) H(\phi)}2 \Big] \;,
$$
whose bosonic piece is
\be
\label{hlocal}
\cL^\text{def}_\text{H} \Big|_\text{bos} = -\Tr \Big[ \big( \sin \theta (D_1\sigma) + \cos \theta Y_3 + iD \big) \, H(\phi) \Big] \;.
\ee
The Gaussian path integral over $D$ imposes
\be
D = i H(\phi) \;.
\ee
Then one is left with
\begin{multline}
\cL^\text{def}_\text{YM} + \cL^\text{def}_\text{H} \Big|_\text{$D$, bos} = \frac12 \Tr \bigg[ \big( D_1 \sigma \cos \theta - Y_3 \sin \theta \big)^2 + \big( H(\phi) - D_1 \sigma \sin \theta - Y_3 \cos \theta \big)^2  \\
+ \big( Y_2 \big)^2 + \big( D_2\sigma \big)^2 + \big( Y_1 \big)^2 + \big( D_3\sigma \big)^2 \bigg] \;,
\end{multline}
which is a sum of squares. The BPS equations are then
\bea
\label{BPSeqnsYMdeformed}
0 &= D_1 \sigma \cos \theta - \Big( F_{12} + \frac\sigma r \Big) \sin \theta \;,\qquad\qquad & 0 &= D_2 \sigma = D_3\sigma \\
0 &= H(\phi) - D_1 \sigma \sin \theta - \Big( F_{12} + \frac\sigma r \Big) \cos \theta \;,\qquad\qquad & 0 &= F_{13} = F_{23} \;.
\eea

Consider now the chiral multiplets, transforming in some representation $\fR = \bigoplus_j \cR_j$ of the gauge and flavor group, where $\cR_j$ are irreducible gauge representations. Imposing the reality conditions $\bar{\phi}^\dagger = \phi$, $\bar{F}^\dagger = F$ and $\sigma^{\dagger} = \sigma$, one finds the BPS equations
\bea
\label{chiraleqn}
0 &= \sin\frac\theta2\, D_+ \phi + \cos\frac\theta2\, \Big( \frac{D_3 + D_3^\dag}2 \phi + \frac qr \phi + \sigma\phi \Big) \;,\qquad\qquad & 0 &= (D_3 - D_3^\dag) \phi \\
0 &= \cos\frac\theta2\, D_- \phi - \sin\frac\theta2\, \Big( \frac{D_3 + D_3^\dag}2 \phi + \frac qr \phi - \sigma\phi \Big) \;,\qquad\qquad & 0 &= F \;,
\eea
where $D_\pm \,\equiv\, D_1 \mp i D_2$ and $D_3 \phi = D_\tau \phi = \big( \partial_\tau - i \frac{a}{2\xi r} -  \frac{1}{2 r} q  -i \frac{\fz}{2\xi r} \big) \phi$.

As before, these equations can be obtained from the canonical deformation action $\cL^\text{def}_\text{mat}$. Its bosonic part reads
\begin{multline}
\cL^\text{def}_\text{mat} \Big|_{\text{bos}} = \frac12 |F|^2 + \frac18 \Big| D_3\phi - D_3^\dagger\phi \Big|^2 + \frac12 \bigg| \sin\frac\theta2\, D_+ \phi + \cos\frac\theta2\, \Big( \frac{D_3 + D_3^\dag}2 \phi + \frac qr \phi + \sigma\phi \Big) \bigg|^2 \\
+ \frac12 \bigg| \cos\frac\theta2\, D_- \phi - \sin\frac\theta2\, \Big( \frac{D_3 + D_3^\dag}2 \phi + \frac qr \phi - \sigma\phi \Big) \bigg|^2  \;.
\end{multline}

\subsection{BPS solutions: Coulomb, Higgs and vortices}
\label{sec: BPS solutions S2S1}

We will now present the BPS solutions to the equations \eqref{BPSeqnsYM}, \eqref{BPSeqnsYMdeformed} and \eqref{chiraleqn}. First, let us recall the solutions for the standard choice $H(\phi) = 0$.

\paragraph{Coulomb-like solutions.}
Consider \eqref{BPSeqnsYM} and \eqref{chiraleqn}. They allow for a field strength
\be
F = \frac{\fm}{2}\sin\theta\,  d\theta \wedge d\varphi \;,
\ee
where $\fm$ can be diagonalized to lie in the Cartan subalgebra and it takes values in the coweight lattice of the gauge group $G$ (it is GNO quantized). The gauge field can be written as
\be
\label{gaugefield}
A=\frac{\fm}{2} \, \big( \kappa - \cos\theta \big)\, d\varphi + \frac{a}{2\xi r} \, d\tau \;,
\ee
where in this section $\kappa = 1$ ($\kappa = -1$) on the patch excluding the south (north) pole. We have also included a holonomy $a$, with $[a,\fm]=0$, around the temporal circle. The BPS equations fix $\sigma = - \fm/2r$ and $D=0$, which is the localization locus of \cite{Imamura:2011su}.

Let us now analyze the BPS equations for a chiral multiplet in gauge representation $\cR$, assuming that its R-charge $q$ is positive, and show that the only smooth solution is $\phi = 0$. First, we decompose $\phi$ in Fourier modes recalling that, in the presence of non-trivial flux on $S^2$, $\phi$ is a section of a non-trivial bundle and should be expanded in monopole harmonics \cite{Wu:1976ge}:
\be
\label{modeexpansion}
\phi(\tau, \theta,\varphi) = \sum_{p,l,m} \; c_{p,l,m} \; \exp\Big( \frac{2\pi i p\tau}{2\xi r} \Big) \; Y_{\frac{\fm}{2},l,m}
\ee
where the range of parameters is $p \in \mathbb{Z}$,  $l \in \frac{|\fm|}{2} + \mathbb{N}$ and $m=-l, -l+1 ,\ldots, +l$. The third component of the angular momentum is given by the eigenvalue of%
\footnote{\label{footnote Wu and Yang}
From \cite{Wu:1976ge}, the gauge invariant angular momentum operator on $\mathbb{R}^3$ in a monopole background $\fm$ is given by $\vec{L} = \vec{r} \times (-i\vec{D}) - \hat{r} \, \fm/2$, where $\hat{r}$ is a unit vector along the $\vec{r}$ direction. In particular the third component is
\be
j_3 =  -i D_\varphi - \fm/2 \cos\theta = -i\partial_\varphi -\kappa \frac{\fm}{2} \;.
\ee
This result is directly applicable to $S^2$. For later reference, we also write the operators $j_+$ and $j_-$:
\be
j_\pm = e^{\pm i\varphi} \Big( \pm \partial_\theta + i \cot \theta\, D_\varphi - \frac{\fm}{2} \sin\theta \Big) = e^{\pm i\varphi} \Big( \pm \partial_\theta + i \cot \theta\, \partial_\varphi + \frac{\fm}{2}\kappa \cot \theta - \frac{\fm}{2} \frac{1}{\sin \theta} \Big) \;.
\ee
}
\begin{align}
j_3 = - i \partial_\varphi - \kappa \frac{\fm}{2},
\end{align}
and on the monopole harmonics: $j_3\  Y_{\frac{\fm}{2},l,m} = m\  Y_{\frac{\fm}{2},l,m}$. Imposing a Hermiticity condition on the holonomy $a$, the equation $(D_3 - D_3^\dag)\phi=0$ corresponds to
\be
\Big( \partial_\tau - i \frac{a}{2\xi r} - i \frac{\mathfrak{z}}{2\xi r}  \Big)\, \phi = 0 \;.
\ee
This implies that only those modes for which $\big( a -2\pi p + \fz \big) \, \phi= 0$ can survive. Since the time dependence is completely fixed, we can reabsorb $p$ by a large gauge transformation and set $p=0$.
From the equations in the first column of \eqref{chiraleqn}, the expressions for $\sigma$ and the gauge field found above, we find $\big( j_3 + \frac q2 \big)\phi =0$ and $j_+\phi = 0$. The first one imposes $m = - q/2$, whereas the second one imposes that the angular momentum eigenvalue $m$ take its maximal value $+l$. For positive R-charge $q>0$, there are no solutions. For zero R-charge (then $l=m=0$) one finds the constant Higgs-like solution $\phi = \phi_0$, if $\big( a + \fz \big) \, \phi= 0$.

\

Now let us see the new solutions with non-trivial $H(\phi)$. We integrate $D$ out first, \ie{} we set $D=iH(\phi)$, solve (\ref{BPSeqnsYMdeformed}) and (\ref{chiraleqn}), and take all vanishing R-charges $q=0$ (arbitrary R-charges can be recovered by analytic continuation of the result by complexifying flavor fugacities). We take exactly the same deformation function $H(\phi)$ as in (\ref{H function}). We find the following classes of solutions.

\paragraph{Deformed Coulomb branch.} It is characterized by $\phi = 0$, and (in complete analogy with \cite{Benini:2012ui}) can be completed to
\be
F_{13} = F_{23} = 0 \;,\qquad F_{12}=2 \zeta \cos\theta + \frac{\fm}{2r^2} \;,\qquad \sigma = - r \zeta \cos \theta - \frac{\fm}{2r} \;.
\ee
We thus have $F_{\theta\varphi} = r^2\sin\theta\, (2 \zeta \cos\theta + \fm/2r^2)$. The corresponding gauge field can be written as
\be
A = \left(r^2 \zeta \sin^2\theta + \frac{\fm}{2}(\kappa - \cos\theta)\right)d\varphi + \frac{a}{2\xi r} d\tau \;.
\ee

\paragraph{Higgs-like solutions.}
They are characterized by $F_{\mu\nu} = 0$, $\sigma = 0$ and a constant profile $\phi$ for the matter fields that solves the D-term equations
\be
H(\phi) = 0 \;,\qquad\qquad\qquad \big( a + \fz \big) \phi = 0 \;.
\ee
The solutions to these algebraic equations are analogous to the Higgs-like solutions of section \ref{sec: solutionsS3}. We will be mainly interested in gauge groups and matter representations such that, for $\zeta_a$ in a suitable range, all VEVs $\phi$ completely break the gauge group.

\paragraph{Vortices.}
Each Higgs-like solution is accompanied by a tower of vortex-string solutions with arbitrary
numbers of vortices at the north and at the south circles. To see this, we expand the BPS equations around $\theta = 0$ and $\theta = \pi$.

The $S^2\times S^1$ metric \eqref{metricS2S1} in the $\theta\to 0$ limit becomes $ds^2 = dR^2 + R^2 d\varphi^2 + d\tau^2$, where $R\equiv r\theta$, which is the metric of $\mathbb{R}^2\times S^1$. The equations \eqref{BPSeqnsYMdeformed} become, to linear order in $R$:
\bea
0 &= D_R \sigma -\frac1r F_{R\varphi} \;,\qquad\qquad & 0 &= D_\varphi \sigma = D_\tau \sigma \\
0 &= H(\phi) - \frac1R F_{R\varphi} - D_R \Big( \frac{\sigma R}{r} \Big) \;,\qquad\qquad & 0 &= F_{\varphi \tau} = F_{R\tau} \;,
\eea
whereas the equations for the chiral fields \eqref{chiraleqn} become
\bea
0 &= \Big( D_R + \frac iR D_\varphi + \frac Rr \sigma \Big) \, \phi \;,\qquad\qquad & 0 &= \big( D_3-D_3^\dagger \big) \phi \\
0 &= \Big( -\frac ir D_\varphi + \sigma  + \frac{D_3 + D_3^\dagger}2 \Big) \, \phi \;,\qquad\qquad & 0 &= F \;.
\eea
Let us qualitatively describe the solutions for a $U(1)$ theory with a single chiral field of charge 1. Working in the gauge $A_\theta = 0$, (\ref{BPSeqnsYMdeformed}) implies that $\partial_\theta \big( r\sigma\cos\theta - A_\varphi \big) = 0$ exactly. We write $r\sigma\cos\theta = A_\varphi - n$, for some integration constant $n$, so it is sufficient to specify the behavior of $\phi$ and $A_\varphi$. Far from the core (the length scale is set by $\sqrt{\zeta^{-1}}$) one finds
\be\label{closetocoreS2S1}
\phi \,\simeq\, \sqrt{\zeta}\, e^{i n \varphi} \;,\qquad\qquad A_\varphi \,\simeq\, n \;,
\ee
and Stokes' theorem implies that $\frac{1}{2\pi}\int F = n$, which is the vortex number. Close to the core:
\be
\phi \,\simeq\, B \, \big( R e^{i  \varphi} \big)^n \;,\qquad\qquad A_\varphi \,\simeq\, 0 + \mathcal{O} \Big( e^{-\frac{R^2}{2r^2}} \Big) \;,
\ee
and in particular $n \geq0$.
A similar analysis can be performed around the south pole in the coordinate $\tilde R =r( \pi - \theta)$. This time we write $r\sigma\cos\theta = A_\varphi - m$. Then $A_\varphi \to 0$ near the core and $A_\varphi \to m$, which we identify with the vortex number, far from the core. We also find $|\phi|\to B' \tilde R^m$ near the core, while it sits in the vacuum far from it: $|\phi|^2 \to \zeta$. The vortex configurations we wrote around the north and south poles are connected by a gauge transformation on the equator: $\phi^N =  e^{i(n-m)\varphi}  \phi^S$ and $A_\varphi^N - A_\varphi^S = n-m$.

For finite values of $\zeta$, we can derive a bound on the allowed vortex numbers. From (\ref{BPSeqnsYMdeformed}) one deduces $H(\phi)\, r\sin\theta = \partial_\theta \sigma$, which results in the inequality
\be
\label{integralH}
r \sigma(\pi) - r \sigma(0) = \frac1{2\pi} \int H(\phi) \dvol(S^2) \leq \zeta \frac{\vol(S^2)}{2\pi} \;,
\ee
which upon plugging in the values of $\sigma$ found above leads to the bound
\be
\label{boundS2S1}
m + n \;\leq\; \zeta \, \frac{\vol(S^2)}{2\pi} \;.
\ee
As in section \ref{sec: solutionsS3}, we conclude that for finite values of $\zeta$ there is a finite number of vortex/antivortex solutions on $S^2$. When the bound is saturated, the chiral field $\phi$ actually vanishes and the gauge field is as in the deformed Coulomb branch described above. We thus get a similar picture of the structure of solutions as in section \ref{sec: solutionsS3}.

\subsection{Computation of the index}

We will now evaluate the classical action and the one-loop determinants of quadratic fluctuations, and then sum/integrate over the space of BPS configurations.

\subsubsection{One-loop determinants from the index theorem}

As in section \ref{sec: 1-loop dets}, we compute the one-loop determinants on non-trivial backgrounds with the equivariant index theorem, following \cite{Drukker:2012sr}. The localizing supercharge squares to
\be\label{qsquaredS2S1}
\cQ^2 =  -\cL^A_{\partial_\tau} + \frac ir \cL^A_{\partial_\varphi} - \cos\theta\, \sigma - \frac{1}{2 r} R + i \sum_j\frac{\mathfrak{z}_j}{2 \xi r} F_j \;.
\ee
The action of $\cQ^2$ on the worldvolume consists of a free rotation along $S^1$ generated by $\cL_{\partial_\tau}$ and a rotation of $S^2$ generated by $\cL_{\partial_\varphi}$ with fixed points at the north and south poles. The equivariant parameters for the $U(1)_{\partial_\varphi} \times U(1)_R \times U(1)_\text{flavor}^F \times G$ are given by $\varepsilon = \frac ir$, $\hat\varepsilon = - \frac1{2 r}$, $\check \varepsilon_j = i \frac{\fz_j}{2\xi r}$ and $\hat a = i A_\tau + \frac1r A_\varphi - \cos\theta\, \sigma$. In appendix \ref{sec: one-loop det} we compute the one-loop determinants in our conventions. For a chiral multiplet in gauge representation $\cR$ we have
\be\label{S2S1chiraloneloop}
Z_\text{1-loop}^\text{chiral} \;\text{``}=\text{''}\; \prod_{w\in\cR} \; \prod_{n\in\bZ} \; \prod_{k\geq0} \frac{i\pi n - (k+1) \xi +\xi \frac q2 - \xi r \, w(\hat a_S) - \frac i2 \sum \fz_j F_j}{i\pi n + k \xi + \xi \frac q2 - \xi r\, w(\hat a_N) - \frac i2 \sum \fz_j F_j} \;,
\ee
which requires regularization. For the gauge multiplet one has
\be\label{S2S1veconeloop}
Z_\text{1-loop}^\text{vec} = \prod_{\alpha > 0} 2 \sinh \big( \xi r \, \alpha(\hat a_N) \big) \, 2 \sinh \big( -\xi r \, \alpha(\hat a_S) \big) \;.
\ee

\subsubsection{Coulomb branch}
\label{sec: Coulomb branch index}

Coulomb branch localization for the 3d index was first performed in \cite{Kim:2009wb} for $\cN=6$ Chern-Simons-matter theories, and later generalized to $\cN=2$ theories in \cite{Imamura:2011su}. A subtlety involving the fermion number was pointed out in \cite{Dimofte:2011ju} (see also \cite{Aharony:2013dha}), and was later confirmed in \cite{Drukker:2012sr} by computing the one-loop determinants with the index theorem. Let us quickly review these results. The Chern-Simons action evaluated on the Coulomb branch configurations gives%
\footnote{We recall that in order to correctly evaluate the CS action $\int A\wedge F$, one should construct an extension $\tilde F$ of the gauge bundle to $S^2 \times D_2$ (where the second factor is a disk) and integrate $\int \tilde F \wedge \tilde F$.}
\be
S_\text{cl}^{CS} =  -\frac{i}{4\pi} \int \Tr_{CS} A\wedge F  = - i \Tr_{CS} a\, \fm \;.
\ee
Due to the modified fermion number, an extra phase $(-1)^{\Tr_{CS} \fm}$ needs to be taken into account \cite{Aharony:2013dha}.

To each Abelian factor (with field strength $F$) in the gauge group is associated a topological symmetry $U(1)_J$, whose current is $J = *F$. Coupling $U(1)_J$ to an external vector multiplet with bosonic components $(A_{BG}, \sigma_{BG}, D_{BG})$ is equivalent to introducing a mixed supersymmetric Chern-Simons term, whose bosonic part is
\be
\label{mixed CS term}
S_J \Big|_\text{bos} = \frac{i}{2\pi} \int \Tr \big( A_{BG} \wedge F + \sigma D_{BG} + \sigma_{BG} D \big) \;.
\ee
An expectation value for $\sigma_{BG}$ would correspond to an FI term. In this section, though, we will be interested in turning on a holonomy $b$ and a flux $\fn$. Notice that this is indeed an example of an external flux for a flavor symmetry, in the spirit of \cite{Kapustin:2011jm}. Evaluation on the Coulomb branch BPS configurations yields
\be
S_J = i \Tr \big( a\, \fn + b\, \fm \big) \;.
\ee
We will introduce the topological fugacity $w=e^{-ib}$. Also in this case extra signs are required: this can be done by taking the index not to be a function of $w$, but rather of $(-1)^{\fn} w$. Such dependence will always be understood.

The gauge equivariant parameter is $\hat a = \dfrac{ia}{2\xi r} + \dfrac\kappa{2 r} \fm$, where $\kappa = 1$ ($-1$) on the northern (southern) patch, as in (\ref{gaugefield}). The chiral one-loop determinant then simplifies and, after regularization, becomes
\be
\label{1-loop Coulomb index}
Z_\text{1-loop}^\text{chiral} = \prod_{w\in\cR} \Big( x^{1-q} \, e^{-i w( a)} \, \zeta^{-F} \Big)^{- w(\fm)/2}       \; \frac{ \big( x^{2 - q - w(\fm)} \, e^{- i w(a)} \, \zeta^{-F} ; x^2 \big)_\infty}{\big( x^{q - w(\fm)} \, e^{i w(a)} \, \zeta^F ; x^2 \big)_\infty} \;,
\ee
where $(a;q)_\infty \equiv \prod_{k=0}^\infty (1-aq^k)$ is the $q$-Pochhammer symbol, we defined $x = e^{-\xi}$ and $\zeta_j = e^{i \fz_j}$, we used the short-hand notation $\zeta^F = \prod_i \zeta_i^{F_i}$, and $q$ is the R-charge. The regularization is similar to \cite{Kim:2009wb} (see also \cite{Drukker:2012sr}).
The expression above includes all the correct extra signs. The vector one-loop determinant becomes
\be
\label{rewritegauge1loop}
Z_\text{1-loop}^\text{vec} = \prod_{\alpha > 0} 4 \sinh\Big( \frac12 \alpha \big( ia+\xi \fm \big) \Big) \sinh\Big( - \frac12 \alpha\big( ia-\xi \fm \big) \Big) = \prod_{\alpha\in G} x^{-\frac12 |\alpha( \fm)|} \Big( 1 - x^{|\alpha( \fm)|}e^{i\alpha(a)} \Big) \;.
\ee
The index is thus computed by the matrix integral:
\be
\label{index Coulomb branch}
I \big( x,\zeta_j, (-1)^{\fn}w, \fn \big) = \frac1{|\mathcal{W}|} \sum_{\fm \,\in\, \bZ^{\rank G}} \int \bigg( \prod_{j=1}^{\rank G} \frac{dz_j}{2\pi i z_j} \bigg) \; (-1)^{\Tr_{CS} \fm} \; e^{i \Tr_{CS} a\fm - i\Tr (a \fn + b \fm)} \; Z_\text{1-loop} \;,
\ee
where $|\mathcal{W}|$ is the order of the Weyl group, $z_j = e^{i a_j}$ is the gauge fugacity and the integration contour is counterclockwise along the unit circle.

\subsubsection{Deformed Coulomb branch}

The full Chern Simons action (\ref{CS action SUSY}) and the mixed CS term (\ref{mixed CS term}) evaluated on the deformed Coulomb branch read
\be
S_\text{cl}^{CS} = - i \Tr_{CS} \big( (a  - 2 i r^2 \xi \zeta) \, \fm \big) \;,\qquad\qquad S_{J} = \Tr \big( (a-2 i r^2 \xi \zeta)\, \fn + b\, \fm \big) \;,
\ee
where, in this subsection, $\zeta$ refers to the deformation parameter (\ref{expansion fake FI}). We also need to include the phase $(-1)^{\Tr_{CS} \fm}$. The equivariant parameter is given by
\be
\hat a = i \frac{a - 2 i r^2  \xi   \zeta}{2\xi r} + \frac{\kappa}{2r} \fm \;.
\ee
As in section \ref{sec: def Coulomb branch S3}, we observe that the net effect of the deformation parameter $\zeta$ is an imaginary shift of the integration variable $a \to a - 2 i r^2  \xi   \zeta$, or equivalently $z \equiv e^{ia} \to x^{-2r^2 \zeta}z$. Effectively it modifies the radius of the integration contour; since $|x|<1$, the contour grows for $\zeta>0$ and shrinks for $\zeta<0$.
The effect on the integral is the same as in section \ref{sec: def Coulomb branch S3}: it remains constant, until the contour crosses some pole and the integral jumps. In view of the bound (\ref{boundS2S1}), this happens precisely when new vortex configuration become allowed, providing the missing residue.

In order to obtain an expression of $Z_{S^2\times S^1}$ purely in terms of vortices, we need to suppress the contribution from the deformed Coulomb branch. Heuristically, this can be achieved if there is no pole at the origin or infinity. As we show in appendix \ref{app: poles zero infinity} following  \cite{Hwang:2012jhPUB}, for a $U(N)$ theory with $N_f$ fundamentals, $N_a$ antifundamentals and Chern-Simons level $k$, there is no pole at infinity if $N_f > N_a$ and $|k|\leq \frac{N_f - N_a}2$, thus suppression is obtained by sending $\zeta \to +\infty$; for  $N_f < N_a$ and $|k| \leq \frac{N_a - N_f}2$ there is no pole at the origin, thus suppression is obtained by sending $\zeta \to -\infty$. For $N_f - N_a = k = 0$ there are poles both at the origin and infinity, however the residues vanish.

\subsubsection{Higgs branch and vortices}
\label{sec: VPF index}

For finite values of the deformation parameters $\zeta_a$, additional BPS configurations are present, namely Higgs vacua and vortex solutions, whose (anti)vortex numbers $(m,n)$ are bounded by (\ref{boundS2S1}) (or its multi-dimensional generalization). We determine here their additional contribution to the path integral, besides the deformed Coulomb branch. The discussion is similar to section \ref{sec: VPF S3}, so we will be brief.

The classical actions can be evaluated exactly using $D = iH(\phi)$, the BPS equations (\ref{BPSeqnsYMdeformed}), the knowledge of the flux carried by the vortices and of the corresponding values of $A_\varphi(\theta)$ at $\theta = 0,\pi$, in a gauge $A_\theta = 0$. Recall that the equations determine $\sigma$ exactly in terms of $A_\varphi$, see around (\ref{closetocoreS2S1}). One finds
\be
\label{classactionvortS2S1}
S_\text{cl}^{CS} = -i \Tr_{CS} \Big( (n-m) a + i \xi ( m^2 - n^2) \Big) \;,\qquad S_J = i \Tr \Big[ \fn \big(a - i\xi (m+n) \big) + b (n-m)\Big] \;,
\ee
where $a$ is evaluated on the Higgs branch, $a = -\fz$. Again we need to include the extra phase $(-1)^{\Tr_{CS} (n-m)}$. The one-loop determinants are evaluated with (\ref{S2S1chiraloneloop}) and (\ref{S2S1veconeloop}), using the equivariant parameters
\be
\hat a_N = \frac{ia + 2\xi n}{2\xi r} \;,\qquad\qquad \hat a_S = \frac{ia + 2\xi m}{2\xi r}
\ee
at the north and south poles, where in both cases $a$ is evaluated on its Higgs branch location $a_H$. The one-loop determinants for the $\rank G$ chiral multiplets Higgsing the gauge group should be computed with a residue prescription. Therefore, after a regularization similar to \cite{Drukker:2012sr}, the one-loop determinant for chiral multiplets is
\be
Z_\text{1-loop}^\text{chiral} = \Res_{a \,\to\, a_H} \; \Bigg[ \prod_{w\,\in\, \fR} \bigg( x^{1 + w(m+n)} \, e^{-i w( a)} \, \zeta^{-F(\phi)} \bigg)^{w(m-n)/2} \;\; \frac{ \big( x^{2 + 2w(m)} \, e^{-iw(a)} \, \zeta^{-F(\phi)} ; x^2 \big)_\infty }{ \big( x^{- 2w(n)} \, e^{iw(a)} \, \zeta^{F(\phi)} ; x^2 \big)_\infty} \Bigg] \;.
\ee
Here $F(\phi)$ refers to the chiral multiplets, $\zeta^F = \prod_i \zeta_i^{F_i}$ and we set the R-charges to zero. For the vector one-loop determinant we have
\bea
Z_{\text{1-loop}}^{\text{gauge}} &= \prod_{\alpha > 0} 2 \sinh\Big( \frac{\alpha (ia + 2\xi n)}2 \Big) \, 2\sinh\Big( - \frac{\alpha(ia + 2\xi m)}2  \Big) \\
&= \prod_{\alpha \,\in\, \fg} x^{-\frac{|\alpha(n-m)|}2} \Big( 1- x^{ |\alpha(n-m)| -\alpha(n+m) } \, e^{i\alpha( a)} \Big) \;,
\eea
evaluated on the Higgs branch location. These expressions, for the vortices that satisfy the bound (\ref{boundS2S1}), precisely reproduce the residues of the integrand in (\ref{index Coulomb branch}), which are the jumps of the deformed Coulomb branch contribution as the contour crosses the poles.

\paragraph{Vortex partition function.}
We will now take a suitable limit $\zeta_a \to \pm\infty$, in which the deformed Coulomb branch contribution is suppressed. Then the resummed contribution of all vortex strings is described by the same vortex partition function that we used on $S^3_b.$

Let us compute the partition function in the limit. First, we have a finite number of Higgs vacua. In each vacuum, the components of the holonomy $a_\alpha$ are fixed to some specific (real) values that are functions of the real masses.
The classical actions (\ref{classactionvortS2S1}) provide an overall classical contribution:
\be
S_J = i \Tr \big( \fn\, a \big) \;,
\ee
as well as the weighting factors for vortices and anti-vortices:
\bea
e^{-S_\text{v}} &= \exp\Big[ - \xi\Tr_{CS} m^2 + \Big(-i\Tr_{CS} a\cdot  \, + \Tr \big( -\xi \fn + i b \big) \cdot \Big) m \Big] \\
e^{-S_\text{av}} &= \exp\Big[ \xi \Tr_{CS} n^2 + \Big( i \Tr_{CS} a \cdot \, +\Tr \big(  -\xi\fn -i  b \big) \cdot \Big) n \Big] \;.
\eea
Second, the one-loop determinants for the vector multiplet and the chiral multiplets not acquiring a VEV are as in the Coulomb branch. The $\rank G$ chiral multiplets acquiring VEV bring a residue factor, which in this case is some phase. Finally, the vortex partition function $Z_\text{vortex}$ depends on equivariant parameters for rotations of $\bR^2$ ($\varepsilon$) and flavor rotations ($g$): they are identified---at $\theta=0$ (N) and $\theta = \pi$ (S)---from the $SU(1|1)$ complex of the supercharge $\cQ$ at the poles, \ie{} from $\cQ^2$ in (\ref{qsquaredS2S1}). We find
\be
\varepsilon_N = -2i\xi \;,\qquad g_N = i \Big( a +\sum\nolimits_j \fz_j F_j \Big) \;,\qquad\qquad
\varepsilon_S = 2i\xi \;,\qquad g_S = -i \Big( a + \sum\nolimits_j \fz_j F_j \Big) \;,
\ee
where the minus sign in the south pole parameters with respect to the north pole ones is due to the opposite orientation.

Eventually, Higgs branch localization gives the following expression for the index:
\be
I = \sum_{\text{Higgs vacua}} e^{- i \Tr (\fn\, a)} \; Z'_\text{1-loop} \; Z_\text{v} \; Z_\text{av} \;.
\ee
The (anti)vortex-string contributions are expressed in terms of the 3d vortex partition function:
\bea
Z_\text{v} &= Z_\text{vortex} \bigg( e^{ - \xi\Tr_{CS} \cdot}  \,,\, e^{ -i\Tr_{CS} a \cdot\,  + \Tr ( -\xi \fn + i b) \cdot }  \,,\,  -2i\xi \,,\, i \Big( a + \sum\nolimits_j \fz_j F_j \Big) \bigg) \\
Z_\text{av} &= Z_\text{vortex} \bigg( e^{ \xi\Tr_{CS} \cdot}  \,,\,  e^{ i\Tr_{CS} a\cdot\, + \Tr ( -\xi \fn - i b) \cdot}    \,,\,  2i\xi \,,\, -i \Big( a +\sum\nolimits_j \fz_j F_j \Big) \bigg) \;.
\eea
As in section \ref{sec: VPF S3}, the first two arguments in the vortex partition function are exponentiated linear functions on the gauge algebra, corresponding to the quadratic and linear weights for the vortex numbers. We shall give a concrete example in the next section.

\subsection{Matching with the Coulomb branch integral}
\label{sec: matching S2S1}

We wish to shortly review, in our conventions, that the superconformal index of a $U(N)$ gauge theory with $N_f$ fundamentals, $N_a$ antifundamentals and CS level
\be
|k| \leq \frac{|N_f - N_a|}2 \;,
\ee
(see footnote \ref{foo: max/min chiral}) can be rewritten in a form that matches with the result of Higgs branch localization, as done in \cite{Hwang:2012jh},%
\footnote{See also \cite{Krattenthaler:2011da}, where the factorized form of the index was first observed in the $U(1)$ case.}
and moreover that the very same $Z_\text{vortex}$ as in (\ref{3d VPF}) emerges.

Concretely,
\begin{multline}
\label{index Coulomb integral U(N)}
I^{U(N), N_f, N_a} = \frac{1}{N!} \sum_{\vec\fm \,\in\, \bZ^N} w^{\sum_j \fm_j} \oint  \prod_{j=1}^N \left(\frac{dz_j}{2\pi i z_j} (-z_j)^{k\fm_j} z_j^{-\fn}\right) \prod_{\substack{i,j = 1\\i\neq j}}^N x^{-|\fm_i - \fm_j|/2} \Big( 1 - z_i z_j^{-1}x^{|\fm_i - \fm_j|} \Big) \\
\times \prod_{i=1}^N \prod_{\alpha = 1}^{N_f} \big( x z_i^{-1} \zeta_\alpha \big)^{-\fm_i/2}  \;  \frac{ \big( z_i^{-1} \zeta_\alpha x^{-\fm_i + 2} \,;\, x^2 \big)_\infty }{ \big( z_i \zeta_\alpha^{-1} x^{-\fm_i } \,;\, x^2 \big)_\infty } \; \prod_{\beta = 1}^{N_a} \big( x z_i \tilde\zeta_\beta^{-1} \big)^{\fm_i/2}  \;  \frac{ \big( z_i \tilde\zeta_\beta^{-1} x^{\fm_i + 2} \,;\, x^2 \big)_\infty }{ \big( z_i^{-1} \tilde\zeta_\beta x^{\fm_i } \,;\, x^2 \big)_\infty } \;,
\end{multline}
where $z_j = e^{ia_j}$ and $w=e^{-ib}$. The flavor fugacities $\zeta_\alpha = e^{i\fz_\alpha}$, $\tilde\zeta_\beta = e^{i\tilde\fz_\beta}$ are defined up to a common rescaling, since the flavor symmetry is $SU(N_f)\times SU(N_a)\times U(1)_A$. The integration contour is along the unit circle for $|\tilde\zeta_\beta| < 1 < |\zeta_\alpha|$.
We also introduced the extra sign $(-1)^{k\sum\fm_j}$, as explained in section \ref{sec: Coulomb branch index}. Note that $k+\frac{N_f+N_a}{2}$ is integer if we impose parity anomaly cancelation: this guarantees that the integrand is a single-valued function of $z_j$.

For $N_f > N_a$ there is no pole at infinity (see appendix \ref{app: poles zero infinity}). Moreover, since $ |\tilde\zeta_\beta| < 1 < |\zeta_\alpha|$ and $|x|<1$, only the one-loop determinants of fundamentals have poles outside the unit circle. More precisely, the numerator of the one-loop determinants of fundamentals has zeros at $z_j  = \zeta_{\alpha_j} x^{-\fm_j + 2r_j}$, for all $r_j \geq 1$ and $j=1,\ldots, N$, while the denominator has zeros at $z_j = \zeta_{\alpha_j } x^{\fm_j -2r_j}$ for all $r_j \geq 0$. For $\fm_j\leq 0$ there is no superposition of zeros, while for $\fm_j > 0$ there is superposition and some of them cancel. The net result is that the poles outside the unit circle are located at
\be
z_j = \zeta_{\gamma_j} \, x^{-|\fm_j|-2r_j} \;,\qquad r_j \in \bZ_{\geq0} \;, \qquad \gamma_j = 1,\ldots,N_f \;,\qquad j=1,\ldots, N \;.
\ee
Summing the residues, one obtains:
\bea
\label{indexrewritten}
I &= \frac{1}{N!} \sum_{\vec\gamma \,\in\, (\bZ_{N_f})^N} \; \sum_{\vec{\mu},\, \vec{\nu} \,\in\, \bZ_{\geq0}^N} (-1)^{{-k}\sum_j (\mu_j - \nu_j)} \, w^{\sum_j (\mu_j - \nu_j)} \prod_{i=1}^N \big( \zeta_{\gamma_i}^{-1} x^{\mu_i + \nu_i} \big)^{{-k}(\mu_i - \nu_i)+\fn} \\
&\quad\times \prod_{i\neq j}^N x^{-\frac{1}{2} {\displaystyle|} (\mu_i - \nu_i ) - (\mu_j - \nu_j) {\displaystyle|}} \bigg( 1 -  \frac{\zeta_{\gamma_j}^{-1} x^{\mu_j + \nu_j}}{ \zeta_{\gamma_i}^{-1} x^{\mu_i + \nu_i}}  \, x^{ {\displaystyle|} (\mu_i - \nu_i) - (\mu_j - \nu_j) {\displaystyle|}} \bigg) \; \prod_{i=1}^N \frac{\big( x^{\mu_i + \nu_i+1} \big)^{-(\mu_i - \nu_i)/2}}{(x^{-2};x^{-2})_{\mu_i}(x^{2};x^{2})_{\nu_i}} \\
&\quad\times \prod_{i=1}^N  \; \prod_{\alpha\, (\neq \gamma_i)}^{N_f}\big(  \zeta_\alpha \zeta_{\gamma_i}^{-1} x^{\mu_i + \nu_i+1} \big)^{-(\mu_i - \nu_i)/2} \; \frac{ \big( \zeta_\alpha \zeta_{\gamma_i}^{-1} x^{2 \nu_i + 2} \,;\, x^2 \big)_\infty }{ \big( \zeta_\alpha^{-1} \zeta_{\gamma_i} x^{-2\mu_i } \,;\, x^2 \big)_\infty } \\
&\quad\times \prod_{i=1}^N \prod_{\beta = 1}^{N_a}  \big( \tilde\zeta_\beta^{-1} \zeta_{\gamma_i} x^{-\mu_i - \nu_i +1}  \big)^{(\mu_i - \nu_i)/2} \; \frac{ \big( \tilde\zeta_\beta^{-1} \zeta_{\gamma_i}  x^{ - 2\nu_i + 2} \,;\, x^2 \big)_\infty }{ \big( \tilde\zeta_\beta \zeta_{\gamma_i}^{-1} x^{2\mu_i } \,;\, x^2 \big)_\infty } \;,
\eea
where we decomposed the summation over $\mu_i = r_i +\frac{\fm_i+|\fm_i|}{2}$ and $\nu_i = \mu_i - \fm_i$. The $q$-Pochhammer symbol is $(a;q)_n = \prod_{k=0}^{n-1} (1-q^ka)$.

At this point one can factorize the summation into a factor independent of $\vec{\mu}$ and $\vec{\nu}$, a summation over $\vec{\mu}$ and a summation over $\vec{\nu}$. One observes that each of the two summations over $\vec{\mu}$ and $\vec{\nu}$ vanishes if we choose $\gamma_i = \gamma_j$ for some $i,j$, and on the other hand it is symmetric under permutations of the $\gamma_i$'s. Therefore we can restrict the sum over unordered combinations $\vec\gamma \in C(N,N_f)$ of $N$ out of the $N_f$ flavors, and cancel the $N!$ in the denominator. Finally, rewriting the $q$-Pochhammer symbols in terms of $\sinh$ and using the identity \eqref{identitysinh} one obtains
\be
I = \sum_{\vec\gamma \,\in\, C(N,N_f)} Z_\text{cl}^{(\vec\gamma)} \; Z_\text{1-loop}^{\prime\,(\vec\gamma)} \; Z_\text{v}^{(\vec\gamma)} \; Z_\text{av}^{(\vec\gamma)} \;.
\ee
The classical and one-loop contributions are
\bea
Z_\text{cl}^{(\vec\gamma)} &= \prod_{j \,\in\, \vec\gamma} \zeta_j^{-\fn} \\
Z_\text{1-loop}^{\prime\,(\vec\gamma)} &= \prod_{j\,\in\,\vec\gamma} \; \prod_{\alpha\, (\neq j)}^{N_f} \frac{ \big( \zeta_j^{-1} \zeta_\alpha x^2 \,;\, x^2 \big)_\infty}{\big( \zeta_j \zeta_\alpha^{-1} \,;\, x^2 \big)_\infty} \prod_{\beta=1}^{N_a} \frac{ \big( \zeta_j \tilde\zeta_\beta^{-1} x^2 \,;\, x^2 \big)_\infty}{ \big( \zeta_j^{-1}\tilde\zeta_\beta \,;\, x^2 \big)_\infty} \,\cdot\, \prod_{\substack{i, j \,\in\, \vec\gamma \\ i\neq j}}  2\sinh \Big(\frac{i \fz_i - i \fz_j}2 \Big) \;.
\eea
The vortex and antivortex contribution can be written as
\bea
Z_\text{v}^{(\vec\gamma)} &= Z_\text{vortex}^{(\vec\gamma)}\Big( e^{-\xi k} \,,\, w_\text{v}^{-1}e^{(-i\fz_j k - \xi \fn)} \big|_{j\in\vec\gamma} \,,\, -2 i \xi \,,\, i \fz_\alpha \,,\, i \tilde\fz_\beta \Big) \\[.4em]
Z_\text{av}^{(\vec\gamma)} &= Z_\text{vortex}^{(\vec\gamma)}\Big( e^{\xi k} \,,\, w_\text{av} \,e^{(i\fz_i k - \xi \fn)} \big|_{i\in\vec\gamma} \,,\, 2 i \xi \,,\, -i \fz_\alpha \,,\,  -i \tilde\fz_\beta \Big) \;,
\eea
and the vortex-string partition function turns out to be exactly the same (\ref{3d VPF}) as for the computation on $S^3_b$, namely:
\begin{multline}\nn
Z_\text{vortex}^{(\vec\gamma)}\Big( Q_j \,,\, L_j \,,\, \varepsilon \,,\, a_\alpha \,,\, b_\beta \Big) = \sum_{\vec\mu \,\in\, \bZ_{\geq0}^N} \; \prod_{j \,\in\, \vec\gamma} \; Q_j^{\mu_j^2}  L_j^{\mu_j} \; (-1)^{(N_f - N_a)\mu_j} \\
\times \prod_{\lambda=0}^{\mu_j-1} \frac{\displaystyle \prod\nolimits_{\beta=1}^{N_a} 2i\sinh \frac{a_j - b_\beta + i\varepsilon\lambda}2}{\displaystyle \prod_{l\,\in\,\vec\gamma} 2i\sinh \frac{a_j - a_l + i\varepsilon(\lambda - \mu_l)}2 \; \prod\nolimits_{\alpha\,\not\in\,\vec\gamma}^{N_f} 2i\sinh \frac{a_\alpha - a_j + i\varepsilon(\lambda - \mu_j)}2} \;.
\end{multline}
The fugacity $w$ for the topological charge is rotated by a phase: $w_\text{v} = (-i)^{N_f-N_a} (-1)^{k+N-1}w$, $w_\text{av} = i^{N_f-N_a} (-1)^{k+N-1}w$. The parameters that determine $Z_\text{v}$ and $Z_\text{av}$ in terms of $Z_\text{vortex}$ are exactly as prescribed by our general discussion in section \ref{sec: VPF index}.

\section{Discussion}

In this paper we have extended the Higgs branch localization framework of \cite{Benini:2012ui} to three-dimensional $\cN=2$ R-symmetric theories on $S^3_b$ and $S^2 \times S^1$. We expect the method to work on much more general 3d backgrounds. We also expect a possible further extension to four-dimensional $\cN=1$ theories on manifolds like $S^3 \times S^1$ or $S^2 \times T^2$ (and fibrations thereof) which naturally support vortex-membranes, \ie{} vortices with 2d worldvolume. Even more generally, the method should work for theories with 8 supercharges, for instance in 4 and 5 dimensions. We leave these investigations to future work.

Higgs branch localization expresses the partition function in terms of the (3d version of the) vortex partition function (VPF), which could also be computed in the $\Omega$-background \cite{Nekrasov:2002qd, Nekrasov:2003rj, Shadchin:2006yz}. In fact, the partition function on different geometries---like $S^3_b$ and $S^2 \times S^1$---is controlled by the very same VPF, with different identifications of the parameters. This has been extensively elaborated upon in \cite{Beem:2012mb}.

In the special case of QCD-like theories, \ie{} $U(N)_k$ gauge theories with $N_f$ fundamentals and $N_a$ antifundamentals, we have noticed that the $S^3$ and $S^2\times S^1$ partition functions factorize into VPFs only for $|k|\leq \frac{|N_f-N_a|}2$, a fact that apparently has been overlooked before. It is a natural question to understand factorization beyond such bound.

It might be worth studying more in detail aspects of the 3d VPF. For instance, 3d mirror symmetry maps particles to vortices \cite{Aharony:1997bx} and it would be interesting to understand its action on the VPF. Through the mirror map \cite{Benini:2010uu} between star-shaped quivers and the 3d reduction of class-$S$ theories \cite{Gaiotto:2009we, Gaiotto:2009hg}, this might shed more light on the latter.

Finally, the VPF encodes (equivariant) geometrical information about the Higgs branch of the theory. It might be interesting to investigate how the VPF captures the quantum moduli space \cite{Gaiotto:2009tk, Benini:2009qs, Jafferis:2009th, Benini:2011cma} of Chern-Simons-matter quiver theories arising from M2-branes at Calabi-Yau fourfold singularities.

\section*{Acknowledgements}

The authors would like to thank Ofer Aharony, Takuya Okuda, Sara Pasquetti and Leonardo Rastelli for useful discussions and correspondence. F.B. would also like to thank Dario Martelli and James Sparks for discussion on related material. F.B.'s work is supported in part by DOE grant DE-FG02-92ER-40697. W.P. is supported in part by NSF Grant PHY-0969919.

\appendix

\section{Spinor conventions}
\label{app: spionor conventions}

We use essentially the same conventions as in \cite{Hama:2010av, Hama:2011ea, Benini:2012ui}. In vielbein space we take the gamma matrices $\gamma^a = \big( \begin{smallmatrix} 0 & 1 \\ 1 & 0 \end{smallmatrix} \big) \,,\, \big( \begin{smallmatrix} 0 & -i \\ i & 0 \end{smallmatrix} \big) \,,\, \big( \begin{smallmatrix} 1 & 0 \\ 0 & -1 \end{smallmatrix} \big)$ which do not have definite symmetry: $[ \gamma^1, \gamma^2, \gamma^3]^\trans = [\gamma^1, - \gamma^2, \gamma^3]$.
We take the charge conjugation matrix $C$, defined by $C \gamma^\mu C^{-1} = - \gamma^{\mu \trans}$, as $C = -i \varepsilon_{\alpha\beta} = \gamma_2$ (where $\varepsilon_{12} = \varepsilon^{12} = 1$) so that
\be
C \gamma^\mu C = - \gamma^{\mu\trans} \;,\qquad\qquad C^2 = \unit \;.
\ee
Indeed $C = C^{-1} = C^\dag = - C^\trans = - C^*$.
Since Dirac spinors are in the \rep{2} of $SU(2)$, there are two products we can consider: $\eta^\trans C \epsilon \equiv -i \, \eta^\alpha \varepsilon_{\alpha\beta} \epsilon^\beta$ and $\eta^\dag \epsilon \equiv \eta^*_\alpha \epsilon^\alpha$. When we use the first product, we omit $^\trans C$ (that is we write $\eta \epsilon \equiv \eta^\trans C \epsilon$). The two products are related by charge conjugation: $\epsilon^c \equiv C \epsilon^*$ and $\epsilon^{c\dag} = \epsilon^\trans C$, so that $\eta^\trans C \epsilon = \eta^{c\dag} \epsilon$. Notice that $(\epsilon^c)^c = - \epsilon$ and there are no Majorana spinors.

Barred spinors will simply be independent spinors. Products are constructed as spelled out before: $\bar\epsilon \lambda \equiv \bar\epsilon^\alpha C_{\alpha\beta} \lambda^\beta$, $\bar\epsilon \gamma^\mu \lambda \equiv \bar\epsilon^\alpha (C\gamma^\mu)_{\alpha\beta} \lambda^\beta$, etc\dots
The charge conjugation matrix $C$ is antisymmetric, while $C\gamma^a$ are symmetric and so $C\gamma^\mu$. Since $\gamma^{\mu\nu}$ equals a single gamma matrix or zero, also $C\gamma^{\mu\nu}$ are symmetric. For anticommuting fermions we get:
\be
\bar\epsilon \lambda = \lambda \bar\epsilon \;,\qquad\qquad \bar\epsilon \gamma^\mu \lambda = - \lambda \gamma^\mu \bar\epsilon \;,\qquad\qquad \bar\epsilon \gamma^{\mu\nu} \lambda = - \lambda \gamma^{\mu\nu} \bar\epsilon \;.
\ee
Some useful relations among gamma matrices are:
\bea
\,[\gamma_\mu, \gamma_\nu] &= 2g_{\mu\nu} \;,\qquad \gamma_\mu \gamma_\nu = g_{\mu\nu} + \gamma_{\mu\nu} \;,\qquad \gamma^{\mu\nu} =i \varepsilon^{\mu\nu\rho} \gamma_\rho \;,\qquad \gamma^{\mu\nu} \varepsilon_{\mu\nu\rho} = 2i \gamma_\rho \\
\gamma_\mu \gamma^{\nu\rho} &= i \varepsilon^{\nu\rho\sigma} g_{\mu\sigma} + (\delta^\nu_\mu \delta^\rho_\alpha - \delta^\nu_\alpha \delta^\rho_\mu) \gamma^\alpha \;,\qquad \gamma_{\mu\nu} \gamma^\nu = - \gamma^\nu \gamma_{\mu\nu} = 2 \gamma_\mu \\
\gamma_\mu \gamma^{\nu\rho} \gamma^\mu &= -\gamma^{\nu\rho} \;,\qquad \gamma_\mu \gamma^\nu \gamma^\mu = - \gamma^\nu \;,\qquad \gamma^\mu \gamma_\mu = 3 \;,\qquad \gamma_{\mu\nu} \gamma^{\rho\nu} = -2 \delta^\rho_\mu - \gamma\du{\mu}{\rho} \\
\gamma^{\mu\nu} \gamma_\rho \gamma_\nu &= -2 \delta^\mu_\rho \;,\qquad \gamma^{\mu\nu} \gamma_\rho \gamma_{\mu\nu} = 2\gamma_\rho \;.
\eea
The antisymmetric tensor with flat indices is $\varepsilon^{\hat1 \hat2 \hat3} = \varepsilon_{\hat1\hat2\hat3} = 1$, and the covariant forms with curved indices are $\varepsilon_{\mu\nu\rho} = \sqrt{g}\, \varepsilon_{\hat\mu\hat\nu\hat\rho}$ and $\varepsilon^{\mu\nu\rho} = \frac1{\sqrt g} \varepsilon^{\hat\mu\hat\nu\hat\rho}$.

The Fierz identity for anticommuting 3d Dirac fermions is
\be
(\bar\lambda_1 \lambda_2) \lambda_3 = - \frac12 (\bar\lambda_1 \lambda_3) \lambda_2 - \frac12 (\bar\lambda_1 \gamma^\rho \lambda_3) \, \gamma_\rho \lambda_2 \;.
\ee
Since $\gamma^\alpha$ and $\gamma^{\mu\nu}$ are dual, one finds
\be
(\gamma_{\mu\rho})_{**} (\gamma^\rho)_{**} = (\gamma^\rho)_{**} (\gamma_{\rho\mu})_{**} \;,\qquad\qquad
-2\, (\gamma_\mu)_{**} (\gamma^\mu)_{**} = (\gamma_{\nu\rho})_{**} (\gamma^{\nu\rho})_{**}
\ee
where indices are not contracted. It might also be useful:
\be
-\frac i4 \bar\epsilon \gamma^\rho \gamma^{\mu\nu}\epsilon\, \gamma_\rho \gamma_\nu \cO_\mu \lambda = - \frac i2 \bar\epsilon \epsilon\, \gamma^\mu \cO_\mu \lambda + \frac i2 \bar\epsilon \gamma^\mu \epsilon\, \cO_\mu \lambda + \frac i4 \bar\epsilon \gamma^{\alpha\rho} \epsilon\, \gamma_\rho \cO_\alpha \lambda \;,
\ee
where $\cO_\mu$ is any operator, acting on any field.

\section{Supersymmetric theories on three-manifolds}
\label{app: SUSY and actions}

Following \cite{Hama:2010av, Hama:2011ea}, we write the superconformal transformation rules on the gauge and matter multiplets on a three-dimensional manifold. The manifold is restricted by the requirement that it admits solutions to the usual Killing spinor equations, and that the superalgebra closes. After presenting the supersymmetry variations, in section \ref{commKS} we present the anticommuting supercharges by replacing the anticommuting Killing spinors in $\delta_{\epsilon}$ and $\delta_{\bar{\epsilon}}$ with their commuting counterparts. Lagrangians invariant under the supersymmetry transformations were studied in \cite{Hama:2010av, Hama:2011ea}. Most of them are exact and therefore will not contribute in a localization computation. Notable exceptions are the Chern-Simons and Fayet-Iliopoulos actions. A more systematic analysis of SUSY on three-manifolds has been done in \cite{Klare:2012gn, Closset:2012ru}.

\subsection{The superconformal algebra}

We define the field strength as $F_{\mu\nu} = \partial_\mu A_\nu - \partial_\nu A_\mu - i [A_\mu, A_\nu]$, and the gauge and metric covariant derivative as $D_\mu = \nabla_\mu - i A_\mu$, where $\nabla_\mu$ is the metric-covariant derivative. It follows, for instance, that for an adjoint scalar $\sigma$: $[D_\mu,D_\nu]\sigma = -i[F_{\mu\nu},\sigma]$. We will also turn on a background gauge field $V_\mu$ for $U(1)_R$, therefore
\be
D_\mu = \nabla_\mu - i A_\mu -iV_\mu \;.
\ee

The superconformal transformations of the vector multiplet are
\bea
\label{vector superconformal transform 1st app}
\delta A_\mu &= - \frac i2 (\bar\epsilon \gamma_\mu \lambda - \bar\lambda \gamma_\mu \epsilon) \hspace{5cm} \delta \sigma = \frac12 (\bar\epsilon \lambda - \bar\lambda \epsilon) \\
\delta\lambda &= \frac12 \gamma^{\mu\nu} \epsilon F_{\mu\nu} - D \epsilon + i \gamma^\mu \epsilon D_\mu \sigma + \frac{2i}3 \sigma \gamma^\mu D_\mu \epsilon \\
\delta\bar\lambda &= \frac12 \gamma^{\mu\nu} \bar\epsilon F_{\mu\nu} + D\bar\epsilon - i \gamma^\mu \bar\epsilon D_\mu \sigma - \frac{2i}3 \sigma \gamma^\mu D_\mu \bar\epsilon \\
\delta D &= - \frac i2 \bar\epsilon \gamma^\mu D_\mu \lambda - \frac i2 D_\mu \bar\lambda \gamma^\mu \epsilon + \frac i2 [\bar\epsilon \lambda, \sigma] + \frac i2 [\bar\lambda \epsilon,\sigma] - \frac i6 ( D_\mu \bar\epsilon \gamma^\mu \lambda + \bar\lambda \gamma^\mu D_\mu \epsilon) \;,
\eea
and those of the chiral multiplet are
\bea
\label{chiral superconformal transform 1st app}
\delta\phi &= \bar\epsilon \psi \qquad\qquad &
\delta\psi &= i \gamma^\mu \epsilon\, D_\mu\phi + i \epsilon \sigma \phi + \frac{2iq}3 \gamma^\mu D_\mu \epsilon \, \phi + \bar\epsilon F \\
\delta\bar\phi &= \bar\psi \epsilon \qquad\qquad &
\delta\bar\psi &= i \gamma^\mu \bar\epsilon \, D_\mu\bar\phi + i \bar\epsilon \bar\phi \sigma + \frac{2iq}3 \gamma^\mu D_\mu \bar\epsilon \, \bar\phi + \epsilon \bar F \\
&& \delta F &= \epsilon \big( i \gamma^\mu D_\mu \psi - i \sigma \psi - i \lambda \phi \big) + \frac i3 (2q-1) D_\mu\epsilon \, \gamma^\mu \psi \\
&& \delta \bar F &= \bar\epsilon \big( i \gamma^\mu D_\mu \bar\psi - i \bar\psi \sigma + i \bar\phi \bar\lambda \big) + \frac i3 (2q-1) D_\mu \bar\epsilon \, \gamma^\mu \bar\psi \;.
\eea
Here $\epsilon$ and $\bar{\epsilon}$ are independent spinors satisfying the Killing spinor equations
\be
D_\mu \epsilon = \gamma_\mu \hat\epsilon \;,\qquad\qquad D_\mu \bar\epsilon = \gamma_\mu \hat{\bar\epsilon} \;,
\ee
in terms of some other spinors $\hat\epsilon, \hat{\bar\epsilon}$. Closure of the algebra requires the additional constraints:
\be
\gamma^\mu \gamma^\nu D_\mu D_\nu \epsilon = -\frac38 \big(R - 2 i V_{\mu\nu} \gamma^{\mu\nu}\big) \, \epsilon \;,\qquad\qquad \gamma^\mu \gamma^\nu D_\mu D_\nu \bar\epsilon = -\frac38 \big(R + 2 i V_{\mu\nu} \gamma^{\mu\nu}\big) \, \bar\epsilon
\ee
with the \emph{same} functions $R$ and $V_{\mu\nu}$ \cite{Hama:2010av,Hama:2011ea}. Consistency implies that $R$ is the scalar curvature of the three-manifold and $V_{\mu\nu} = \partial_\mu V_\nu - \partial_\nu V_\mu$ is the background gauge field strength. Then the algebra reads
\be
\label{SUSY commutators}
[\delta_\epsilon, \delta_{\bar\epsilon}] = \cL^A_\xi  + i\Lambda +  \rho \Delta + i\alpha R, \qquad\qquad [\delta_\epsilon, \delta_{\epsilon}]=0, \qquad\qquad [\delta_{\bar\epsilon}, \delta_{\bar\epsilon}]=0 \;,
\ee
where $\cL^A_\xi$ is the gauge-covariant Lie derivative (independent of the metric, see below) along the vector field $\xi$, $i\Lambda$ denotes a gauge transformation with parameter $i\Lambda$, $R$ is the R-symmetry charge,%
\footnote{The R-charges are:
\be
R(A_\mu,\sigma,\lambda,\bar{\lambda},D) = (0,0, -1, 1, 0) \;,\quad R(\phi,\bar\phi,\psi,\bar\psi,F,\bar F)=(q,-q,q-1,1-q,q-2,2-q) \;,\quad R(\epsilon, \bar\epsilon) = (-1,+1) \;.
\ee}
and $\Delta$ the scaling weight.%
\footnote{The dilation weights are:
\be
\Delta(A_\mu,\sigma,\lambda,\bar{\lambda},D) = (1,1, \tfrac32, \tfrac32, 2) \;,\quad \Delta(\phi,\bar\phi,\psi,\bar\psi,F,\bar F) = (q,q,q + \tfrac12, q + \tfrac12,q+1,q+1) \;,\quad \Delta(\epsilon, \bar\epsilon) = (\tfrac12,\tfrac12) \;.
\ee
Note that whereas the 1-form $A$ has weight zero, its components have weight 1. The commutator on $A_\mu$ gives the $\mu$-component of the Lie derivative on the 1-form $A$, without further action of the dilation group. }
The parameters themselves are given by
\bea
\label{def of bilinears}
\xi^\mu &= i \bar\epsilon \gamma^\mu \epsilon \qquad\qquad &
\rho &= \frac i3 ( D_\mu \bar\epsilon \gamma^\mu \epsilon + \bar\epsilon \gamma^\mu D_\mu \epsilon) = \frac13 D_\mu \xi^\mu \\
\Lambda &= \bar\epsilon \epsilon \sigma \qquad\qquad &
\alpha &= - \frac 13 ( D_\mu \bar\epsilon \gamma^\mu \epsilon - \bar\epsilon \gamma^\mu D_\mu \epsilon) - \xi^\mu V_\mu \;.
\eea

The Lie derivative $\cL_X$ with respect to a vector field $X$ is a derivation independent of the metric. On forms it is easily defined as $\cL_X = \{d, \iota_X\}$ in terms of the contraction $\iota_X$; using the normalization $\alpha = \frac1{n!} \alpha_{\mu_1 \cdots \mu_n} dx^{\mu_1 \cdots \mu_n}$, in components we have
\be
[\cL_X \alpha]_{\mu_1 \cdots \mu_n} = X^\mu \partial_\mu \alpha_{\mu_1 \cdots \mu_n} + n \, (\partial_{[\mu_1} X^\mu) \, \alpha_{\mu | \mu_2 \cdots \mu_n]} \;.
\ee
The Lie derivative of spinors \cite{Kosmann:1972} (see \cite{Godina:2003tc} for explanations) is
\be
\cL_X \psi = X^\mu \nabla_\mu \psi + \frac14 \nabla_\mu X_\nu \, \gamma^{\mu\nu} \psi \;,
\ee
where the covariant derivative is $\nabla_\mu = \partial_\mu + \frac14 \omega_\mu^{ab} \gamma_{ab}$. Although this definition seems to depend on the metric (through the spin connection and the vielbein), the dependence in fact cancels out. Finally, we can define a ``gauge-covariant'' Lie derivative that acts on sections of some (gauge) vector bundle. On tensors it is simply obtained by substituting the flat derivative $\partial_\mu$ with the covariant derivative, $\partial_\mu \to \partial^A_\mu = \partial_\mu - i A_\mu$, while on spinors it is obtained by substituting $\nabla_\mu \to \nabla_\mu^A$ in the first term. The gauge-covariant Lie derivative of the connection (which does not transform as a section of the adjoint bundle) is defined as
\be
\cL_X^A A = \cL_X A - d^A(\iota_X A) \;,\qquad\qquad (\cL_X^A A)_\mu = X^\rho F_{\rho\mu} = X^\rho \big( 2\partial_{[\rho} A_{\mu]} - i[A_\rho, A_\mu] \big) \;.
\ee

\subsection{Commuting Killing spinors}
\label{commKS}

For given anticommuting spinors $\epsilon, \bar\epsilon$, let us construct the corresponding supercharges $Q,\tilde Q$ in terms of \emph{commuting} spinors $\epsilon$ and $\tilde\epsilon = -C\bar\epsilon^*$ (so that $\bar\epsilon = \tilde\epsilon^c$). They are constructed as follows:
\be
\delta = \delta_\epsilon + \delta_{\bar\epsilon} = \epsilon^\alpha Q_\alpha + \bar\epsilon^\alpha \tilde Q_\alpha \;,\qquad\qquad
Q = \epsilon^\alpha Q_\alpha \;,\qquad\qquad \tilde Q = \tilde\epsilon^{c\, \alpha} \tilde Q_\alpha = - (\tilde\epsilon^\dag C)^\alpha \tilde Q_\alpha \;.
\ee
We also need the charge conjugate $\bar\lambda = C (\lambda^\dag)^\trans$. On the vector multiplet we get:
\bea
\label{gaugemultiplet}
Q A_\mu &= \frac i2 \lambda^\dag \gamma_\mu \epsilon \hspace{2cm}
Q\lambda = \frac12 \gamma^{\mu\nu} \epsilon F_{\mu\nu} - D\epsilon + i \gamma^\mu \epsilon \, D_\mu\sigma + \frac{2i}3 \sigma \gamma^\mu D_\mu\epsilon \hspace{-3.5cm} \\
\tilde Q A_\mu &= \frac i2 \tilde\epsilon^\dag \gamma_\mu \lambda  \hspace{2cm}
\tilde Q \lambda^\dag = - \frac12 \tilde\epsilon^\dag \gamma^{\mu\nu} F_{\mu\nu} + \tilde\epsilon^\dag D + i \tilde\epsilon^\dag \gamma^\mu D_\mu\sigma + \frac{2i}3 D_\mu \tilde\epsilon^\dag \gamma^\mu \sigma \hspace{-6cm} \\
QD &= - \frac i2 D_\mu \lambda^\dag \gamma^\mu \epsilon + \frac i2 [\lambda^\dag \epsilon, \sigma ] - \frac i6 \lambda^\dag \gamma^\mu D_\mu\epsilon &
\tilde Q \lambda &= 0 \qquad &
Q\sigma &= - \frac12 \lambda^\dag \epsilon \\
\tilde Q D &= \frac i2 \tilde\epsilon^\dag \gamma^\mu D_\mu \lambda + \frac i2 [\sigma, \tilde\epsilon^\dag\lambda] + \frac i6 D_\mu \tilde\epsilon^\dag \gamma^\mu \lambda &
Q \lambda^\dag &= 0 &
\tilde Q \sigma &= - \frac12 \tilde\epsilon^\dag \lambda \;.
\eea
On the chiral multiplet we get:
\bea
\label{chiralmultiplet}
Q\phi &= 0 &
\tilde Q \phi &= - \tilde\epsilon^\dag \psi \\
Q\phi^\dag &= \psi^\dag \epsilon &
\tilde Q \phi^\dag &= 0 \\
Q\psi &= \big( i \gamma^\mu D_\mu \phi + i\sigma \phi) \epsilon + \frac{2iq}3 \phi \, \gamma^\mu D_\mu \epsilon &
\tilde Q \psi &= C \tilde\epsilon^* F \\
\tilde Q \psi^\dag &= \tilde\epsilon^\dag \big( -i\gamma^\mu D_\mu \phi^\dag + i \phi^\dag \sigma \big) - \frac{2iq}3 D_\mu \tilde\epsilon^\dag \gamma^\mu \phi^\dag &
Q \psi^\dag &= - \epsilon^\trans C F^\dag \\
QF &= \epsilon^\trans C \big( i \gamma^\mu D_\mu\psi - i \sigma\psi - i \lambda\phi \big) + \frac{i(2q-1)}3 D_\mu \epsilon^\trans C \gamma^\mu \psi &
\tilde Q F &= 0 \\
\tilde Q F^\dag &= \big( - i D_\mu \psi^\dag \gamma^\mu - i \psi^\dag \sigma + i \phi^\dag \lambda^\dag \big) C \tilde\epsilon^* - \frac{i(2q-1)}3 \psi^\dag \gamma^\mu C D_\mu \tilde\epsilon^* \qquad &
Q F^\dag &= 0 \;.
\eea
Finally we define $\mathcal{Q} \equiv Q + \tilde{Q}$.

\subsection{Supersymmetric actions}
\label{app: SUSY actions}

Let us write down the $\cQ$-closed but not $\cQ$-exact actions we consider in the paper: they are the Chern-Simons (CS) action and the Fayet-Iliopoulos (FI) action. Since they are non-trivial in $\cQ$-cohomology, their evaluation on the BPS configurations is non-trivial. The CS action is
\be
\label{CS action SUSY}
S_\text{CS} = - \frac{i}{4\pi} \int \Tr_{CS} \bigg[ A \wedge F - \frac{2i}3 A \wedge A \wedge A + \Big( 2D\sigma - \bar\lambda\lambda \Big) \dvol \bigg] \;,
\ee
both on $S^3_b$ and $S^2 \times S^1$. The symbol $\Tr_{CS}$ (as in \cite{Kapustin:2009kz}) means a trace where each Abelian and simple factor in the gauge group is weighed by its own (quantized) CS level $k$. For instance, for $SU(N)$ this would just be $\Tr_{CS} = k \Tr$.

The FI action on $S^3_b$ is
\be
S_\text{FI} = \frac{i}{2\pi \sqrt{\ell\tilde\ell}} \int \Tr_{FI} \Big( D - \frac\sigma f\Big) \, \dvol(S^3_b) \;,
\ee
where again $\Tr_{FI}$ is a trace where each Abelian factor is weighed by its own FI term $\xi$. For $U(N)$, this would just be $\Tr_{FI} = \xi \Tr$.

\section{One-loop determinants from an index theorem}
\label{sec: one-loop det}

The one-loop determinants of quadratic fluctuations around a non-trivial background, in particular around our general vortex backgrounds, are most easily evaluated with the help of an equivariant index theorem for transversally elliptic operators \cite{Atiyah:1974}. Such a technique was used on $S^4$ \cite{Pestun:2007rz, Gomis:2011pf} and $S^2$ \cite{Benini:2012ui}, while the computations on $S^3_b$ and $S^2\times S^1$ have been done in \cite{Drukker:2012sr}. We will summarize the latter computation here, adapted to our conventions, referring to \cite{Pestun:2007rz, Gomis:2011pf, Benini:2012ui, Drukker:2012sr} for details.

After the cancelations between bosons and fermions, the one-loop determinant equals the ratio $\det_{\text{coker }D_{oe}} \cQ^2/\det_{\text{ker } D_{oe}} \cQ^2$, where $D_{oe}$ is the projection, from a subset $\{\varphi_e\}$ to a subset $\{\varphi_o\}$ of fields, of the expansion of $\cQ$ at linear order around the background. The ratio of weights of the group action of $\cQ^2$ on respective spaces can be computed by first evaluating the index
\be
\ind D_{oe}(\epsilon) = \tr_{\text{ker }D_{oe}} e^{\cQ^2(\epsilon)} - \tr_{\text{coker }D_{oe}} e^{\cQ^2(\epsilon)} \;,
\ee
where $\epsilon$ summarizes the equivariant parameters, and then extracting the determinant with the map
\be
\label{index map}
\sum\nolimits_\alpha c_\alpha e^{w_\alpha(\epsilon)} \quad\to\quad \prod\nolimits_\alpha w_\alpha(\epsilon)^{c_\alpha} \;.
\ee

As explained in the main text, $\ind D_{oe}(\epsilon)$ is computed with the help of the index theorem, and it only gets contributions from the fixed points on the worldvolume of the action of $\cQ^2$. However the theorem can be applied if the action is compact, which is not the case on $S^3_b$ and $S^2 \times S^1$ in general. Then \cite{Drukker:2012sr} propose to reduce along an $S^1$ fiber, and be left with the computation on $S^2$, as in \cite{Benini:2012ui}. It turns out that for the chiral multiplet the operator $D_{oe}$ is the Dolbeault operator $D_{\bar z}$ with inverted grading acting on $\Omega^{(0,0)}$, whose index is $- \frac1{1-z}$, while for the vector multiplet it is the real operator $d^* \oplus d$ acting on $\Omega^1$, whose index is $\frac12$.

\paragraph{The sphere $\boldsymbol{S^3_b}$.} We write the metric in Hopf coordinates as in (\ref{Hopf coordinates}), in terms of $\phi_H = \varphi - \chi$ and $\psi_H = \varphi + \chi$. The square of the supercharge is
\bea
\cQ^2 &= \cL_\xi^A - \sigma - \frac i2 \Big( \frac1\ell + \frac1{\tilde\ell} \Big) R = \frac br \cL^A_{\partial_\varphi} + \frac{b^{-1}}r \cL^A_{\partial_\chi} - \frac{r\sigma}r - \frac i{2r} (b+b^{-1}) R \\
&= \frac{b+b^{-1}}r \cL^A_{\psi_H} + \frac{b-b^{-1}}r \cL^A_{\phi_H} - \frac{r\sigma}r - \frac i{2r} (b + b^{-1}) R
\eea
where we used $r = \sqrt{\ell \tilde\ell}$ and $b = \sqrt{\tilde\ell/\ell}$.

At the northern circle, $\theta = 0$, the Hopf fiber is parametrized by $\varphi$ (see (\ref{metricS3b})) and $\cQ^2$ acts freely on it with equivariant parameter $b$; the KK modes thus contribute $\sum_{n\in\bZ} e^{ibn}$ to the index. On the $S^2$, parametrized by $\theta$ and $\phi_H$, resulting from the reduction along the Hopf fiber, $\cQ^2$ has a fixed point at $\theta = 0$. There the SUSY variation of a chiral multiplet (see (\ref{BPS eqns components matter})) is schematically $D_\theta + \frac i\theta D_{\phi_H} \sim D_{\bar z}$ if we identify $z = \theta e^{i\phi_H}$. In fact the one-loop determinant of the chiral multiplet is the index of the Dolbeault operator with inverted grading (as noticed in \cite{Pestun:2007rz, Gomis:2011pf, Benini:2012ui}), which is $- \frac1{1-z}$. Now we expand in $t = e^{i\phi_H}$ and use the equivariant parameter $(b-b^{-1})$, getting $-\sum_{m\geq 0} e^{i(b-b^{-1})m}$. Putting everything together, and recalling that the multiplet transforms in a gauge representation $\cR$, the contribution to the index of a chiral multiplet from the northern circle is:
\be
\text{ind chiral}_N = - \sum_{w\in\cR} \sum_{n \in \bZ} e^{ibn} \sum_{m\geq0} e^{i(b-b^{-1})m}  e^{- \frac i2 QR} e^{w(\hat a_N)}
\ee
where $Q \equiv b + b^{-1}$ and $\hat a = -i \big( b A_\varphi + b^{-1} A\chi \big) - r\fS$.

At the southern circle, $\theta = \frac\pi2$, the Hopf fiber is parametrized by $\chi$ and $\cQ^2$ acts freely on it with equivariant parameter $b^{-1}$, therefore the KK modes yield $\sum_{n\in\bZ} e^{ib^{-1}n}$. The SUSY variation around $\theta = \frac\pi2$ is schematically $-D_{\tilde\theta} + \frac i{\tilde\theta} D_{\phi_H} \sim D_{\bar z}$ (where $\tilde \theta = \frac\pi2 - \theta$) if we identify $z = \tilde\theta e^{-i\phi_H}$. Again we expand in $t$ and use equivariant parameter $(b-b^{-1})$, getting $\sum_{m\geq1} e^{i(b-b^{-1})m}$. Putting together:
\be
\text{ind chiral}_S = \sum_{w\in\cR} \sum_{n \in \bZ} e^{ib^{-1}n} \sum_{m\geq1} e^{i(b-b^{-1})m} e^{-\frac i2 QR} e^{w(\hat a_S)} \;.
\ee
The one-loop determinant is extracted with (\ref{index map}). We get the non-regulated expression
\be
Z_\text{1-loop}^\text{chiral} \text{``}=\text{''} \prod_{w\in\cR} \prod_{n\in\bZ} \prod_{m\geq 0} \frac{(m+1)b + nb^{-1} - \frac Q2 R - iw(\hat a_S)}{nb - mb^{-1} - \frac Q2 R - iw(\hat a_N)} \;.
\ee
This is the expression in (\ref{1loopchiral long}), after a rescaling by $\sqrt{\ell\tilde\ell}$ of both numerator and denominator.
If $\hat a_N \neq \hat a_S$, this expression cannot be further simplified; the regulated expression could be written in terms of infinite $q$-Pochhammer factors. In our case $\hat a_N = \hat a_S \equiv \hat a$, thus we can simplify coincident factors and, neglecting overall signs, we get
\be
Z_\text{1-loop}^\text{chiral} \text{``}=\text{''} \prod_{w\in\cR} \prod_{m,n\geq 0} \frac{mb + nb^{-1} + \big( 1 - \frac R2 \big)Q -iw(\hat a)}{mb + nb^{-1} + \frac R2 Q + i w(\hat a)} = \prod_{w\in\cR} s_b \bigg( \frac{iQ}2 (1-R) + w(\hat a) \bigg)\;.
\ee
This is the expression in (\ref{1loopchiral}), and the last regulated expression was found in \cite{Hama:2011ea} in terms of the double sine function.

The one-loop determinant of the vector multiplet is computed in a similar way, observing that the relevant complex is the de Rham complex: the index of its complexification is just 1, therefore we get $\frac12$. At the northern and southern circles the indices are $\frac12 \sum_{n\in\bZ} e^{ibn + \alpha(\hat a_N)}$ and $\frac12 \sum_{n\in\bZ} e^{ib^{-1}n + \alpha(\hat a_S)}$ respectively, summed over the roots $\alpha$ of the gauge group. Extracting the eigenvalues and regularizing, we get
\be
Z_\text{1-loop}^\text{vec} = \prod_{\alpha > 0} 2 \sinh\big( \pi b^{-1} \, \alpha(\hat a_N) \big) \, 2 \sinh\big( \pi b\, \alpha(\hat a_S) \big) \;,
\ee
where the product is over the positive roots and the normalization is somewhat arbitrary.

\paragraph{The space $\boldsymbol{S^2\times S^1}$.}
The square of the supercharge reads in this case
\be
\mathcal{Q}^2 =  -\mathcal{L}^A_{\partial_\tau} + \frac{ i}{ r} \mathcal{L}^A_{\partial_\varphi} - \cos\theta\, \sigma - \frac{1}{2 r} R + i \frac{\mathfrak{z}_j}{2 \xi r} F_j \;.
\ee
It generates a free rotation along $S^1$ (of radius $2\xi r$) with equivariant parameter $-1$, thus resulting in the KK contribution $\sum_{n\in\bZ} e^{-i\pi n/\xi r}$, and a rotation of the base $S^2$ with fixed points at $\theta=0$ and $\theta=\pi$.

At $\theta = 0$ the SUSY variation of a chiral multiplet is of the form $D_\theta + \frac i\theta D_{\varphi} \sim D_{\bar z}$ if we identify $z = \theta e^{i\varphi}$. As above, the one-loop determinant of the chiral multiplet is then obtained from the index of the Dolbeault operator with inverted grading, which is $- \frac1{1-z}$. We expand in $t = e^{i\varphi}$ and use the equivariant parameter $ \frac{i}{r}$, getting $-\sum_{k\geq 0} e^{-k/r}$. The total index at the north pole is thus:
\be
\text{ind chiral}_N = - \sum_{w\in \mathcal{R}}\sum_{n \in \bZ} e^{-\pi i n/\xi r} \sum_{k\geq0} e^{-k/r}  e^{ - \frac{1}{2 r} R} e^{i \frac{\mathfrak{z}_j}{2 \xi r} F_j} e^{w(\hat a_N)}
\ee
where $\hat a = i A_\tau + \frac{1}{ r} A_\varphi - \cos\theta\, \sigma$. Similarly, at $\theta = \pi$ the SUSY variation is of the form $D_{\tilde \theta} + \frac{i}{\tilde \theta} D_\varphi \sim D_{\bar z}$ (where $\tilde\theta = \pi - \theta$) if we identify $z = \tilde\theta e^{i\varphi}$. Now we expand the index of the Dolbeault operator in $t^{-1}$ (since the orientation is opposite) and use the equivariant parameter $\frac{i}{r}$, getting $\sum_{k \geq 1} e^{k/r}$. The total index at the south pole is thus:
\be
\text{ind chiral}_S =  \sum_{w\in \mathcal{R}}\sum_{n \in \bZ} e^{-\pi i n/\xi r} \sum_{k\geq 1} e^{k/r}  e^{ - \frac{1}{2 r} R} e^{i \frac{\fz_j}{2 \xi r} F_j} e^{w(\hat a_S)} \;.
\ee
The one-loop determinant is extracted with (\ref{index map}), obtaining the non-regulated expression:
\be
Z_\text{1-loop}^\text{chiral} \text{``}=\text{''} \prod_{w\in \mathcal{R}}\prod_{n\in \mathbb{Z}}\prod_{k\geq 0} \frac{  -\pi i n + (k+1)\xi - \frac{\xi}{2} R + \frac i2\sum_j \fz_j F_j + \xi r \, w(\hat a_S) }{ -\pi i n - k\xi - \frac{\xi}{2} R + \frac i2 \sum_j\fz_j F_j + \xi r \, w(\hat a_N ) } \;.
\ee

For the vector multiplet, a computation exactly parallel to the one for $S^3_b$ gives
\bea
Z_\text{1-loop}^\text{vec} \text{``}&=\text{''} \prod_{\alpha \in \fg}\prod_{n\in \mathbb{Z}} \Big( \alpha(\hat a_N) + \frac{2\pi i n}{2\xi r} \Big)^{1/2} \Big( \alpha(\hat a_S) +\frac{2\pi i n}{2\xi r} \Big)^{1/2} \\
&= \prod_{\alpha > 0} 2 \sinh\big( \xi r \, \alpha(\hat a_N) \big) \, 2 \sinh\big( -\xi r \, \alpha(\hat a_S) \big)\;,
\eea
where the product runs over the positive roots.

\section{One-loop deteminants on $S^2 \times S^1$: poles at zero or infinity}
\label{app: poles zero infinity}

In this appendix we study under what conditions the one-loop determinants on $S^2\times S^1$ do not have poles at zero or infinity, and therefore the deformed Coulomb branch contribution can be suppressed in a suitable $\zeta\to\pm\infty$ limit---$\zeta$ being the coefficient in (\ref{H function}) and (\ref{expansion fake FI})---or equivalently the Coulomb branch contribution can be reduced to a sum of residues as in section \ref{sec: matching S2S1}. For simplicity we consider the case of a $U(1)$ gauge theory with $N_f$ fundamentals, $N_a$ antifundamentals and CS level $k$; the case of $U(N)$ gauge group is a straightforward generalization. We follow an argument in \cite{Hwang:2012jhPUB}, correcting a small imprecision.

First, we remind that the chiral one-loop determinant on the Coulomb branch can be written in two ways:
\bea
Z^\text{chiral}_\text{1-loop} &= \prod_{w\in\cR} \Big( x^{1-q} \, e^{-i w( a)} \, \zeta^{-F} \Big)^{- w(\fm)/2} \; \frac{ \big( x^{2 - q - w(\fm)} \, e^{- i w(a)} \, \zeta^{-F} ; x^2 \big)_\infty}{\big( x^{q - w(\fm)} \, e^{i w(a)} \, \zeta^F ; x^2 \big)_\infty} \\
&= \prod_{w\in\cR} (-1)^{\frac{w(\fm) + |w(\fm)|}2} \Big( x^{1-q} \, e^{-i w( a)} \, \zeta^{-F} \Big)^{ |w(\fm)|/2} \; \frac{ \big( x^{2 - q + |w(\fm)|} \, e^{- i w(a)} \, \zeta^{-F} ; x^2 \big)_\infty}{\big( x^{q + |w(\fm)|} \, e^{i w(a)} \, \zeta^F ; x^2 \big)_\infty} \;.
\eea
The first line is as in (\ref{1-loop Coulomb index}), where $\zeta^F \equiv \prod_j \zeta_j^{F_j} = \prod_j e^{i\fz_j F_j}$ are the flavor fugacities; the equality with the second line can be proven easily.

The index of the $U(1)$ theory is then computed by (we set the R-charges $q=0$):
\begin{multline}
I_\infty = \sum_{\fm \,\in\, \bZ} (-1)^{k\fm + N_f\frac{|\fm| + \fm}2 + N_a\frac{|\fm|-\fm}2} \; w^{\fm} \; x^{\frac{N_f+N_a}2|\fm|} \prod_{\alpha = 1}^{N_f} \big( \zeta_\alpha \big)^{|\fm|/2}  \prod_{\beta = 1}^{N_a} \big( \tilde\zeta_\beta^{-1} \big)^{|\fm|/2} \\
\oint \frac{dz}{2\pi i z} \; z^{k\fm-\fn-\frac{1}{2}(N_f-N_a)|\fm|} \; A_\infty(N_f,N_a,x,\zeta,\tilde \zeta,z; \fm)\;,
\end{multline}
where we introduced
\be
\label{defAinf}
A_\infty(N_f,N_a,x,\zeta,\tilde \zeta, z; \fm) = \prod_{\alpha = 1}^{N_f} \frac{ \big( z^{-1} \zeta_\alpha x^{|\fm| + 2} \,;\, x^2 \big)_\infty }{ \big( z \zeta_\alpha^{-1} x^{|\fm| } \,;\, x^2 \big)_\infty } \; \prod_{\beta = 1}^{N_a}  \;  \frac{ \big( z \tilde\zeta_\beta^{-1} x^{|\fm| + 2} \,;\, x^2 \big)_\infty }{ \big( z^{-1} \tilde\zeta_\beta x^{|\fm| } \,;\, x^2 \big)_\infty } \;.
\ee
This is exactly the same integral as in (\ref{index Coulomb integral U(N)}), in the special case $N=1$. In particular $z=e^{ia}$, the integration contour is along the unit circle $|z|=1$ for $|\tilde\zeta_\beta| < 1 < |\zeta_\alpha|$, and convergence of the Pochhammer symbols requires $|x|<1$. For fixed $\zeta_\alpha$, $\tilde\zeta_\beta$, $x$, the product $A_\infty$ is uniformly convergent on the unit circle $|z|=1$ and the convergence is faster the larger is $|\fm|$, therefore one can argue as in \cite{Hwang:2012jhPUB} that for every $\varepsilon$ there is an $n$ such that
\be
\big| A_\infty(N_f,N_a,x,\zeta,\tilde \zeta, z; \fm) - A_n(N_f,N_a,x,\zeta,\tilde \zeta, z; \fm) \big| < \varepsilon \qquad\qquad \forall\; |z|=1 \;,\quad \forall\; \fm \;,
\ee
where $A_n$ is the same quantity as in \eqref{defAinf} with $\infty$ replaced by $n$ in the Pochhammer symbols. Then one can argue that
\be
\big| I_\infty - I_n \big| \;\leq\; \varepsilon \sum_{m \,\in\, \bZ} \Big| x^{N_f + N_a} \prod_{\alpha=1}^{N_f} \zeta_\alpha \prod_{\beta=1}^{N_a} \tilde\zeta^{-1}_\beta \Big|^{|\fm|/2} |w|^\fm \;,
\ee
where the right-hand-side is finite and $\cO(\varepsilon)$ for small enough $x$. We can thus approximate $I_\infty$ arbitrarily well by $I_n$ by choosing a large enough $n$.

To compute $I_n$, we can deform its integration contour either towards infinity or zero and pick up residues. We can rewrite
\be
A_n = z^{-n(N_f - N_a)} \prod_{j=0}^{n-1} \prod_{\alpha=1}^{N_f} \frac{z - \zeta_\alpha x^{|\fm|+2j+2}}{1-z\zeta_\alpha^{-1} x^{|\fm|+2j}} \prod_{\beta=1}^{N_a} \frac{1-z \tilde\zeta_\beta^{-1} x^{|\fm|+2j+2}}{z - \tilde\zeta_\beta x^{|\fm|+2j}} \;.
\ee
The only factor that can contribute poles either at zero or infinity is $z^{-n(N_f-N_a)}$.

For $N_f > N_a$, $A_n(z)$ does not provide poles at infinity for any arbitrarily large $n$ and we can deform the integration contour towards infinity. However the integrand also contains $z^{k\fm - \frac{N_f - N_a}2 |\fm| - \fn}$; we have absence of poles at infinity for all $\fm\in\bZ$ if
\be
|k| \leq \frac{N_f - N_a}2 \;.
\ee
Here it is important to note that, when evaluating $I_n$, $n$ is held fixed while $\fm$ is summed over $\bZ$. Theories with $|k|$ within the bound have been dubbed ``maximally chiral'' in \cite{Benini:2011mf}. If $|k|$ is larger than the bound, $I_n$ receives contributions from poles at infinity for infinitely many values of $\fm$, and such contributions do not disappear in the $n\to\infty$ limit; therefore the mere sum of the residues not at infinity does not reproduce the correct result. We will not attempt to perform the complete computation in this paper.

For $N_f < N_a$, $A_n(z)$ does not have poles at zero and we can deform the integration contour towards zero. Because of the extra factor in the integrand, there are no poles at $z=0$ for all $\fm\in\bZ$ if
\be
|k| \leq \frac{N_a - N_f}2 \;.
\ee

The case $N_f = N_a$ is a bit special, because $A_n(z)$ has a finite non-zero value both at $z=0$ and $z=\infty$. Consider $k=0$. For $\fn\geq 1$ there are no poles at infinity, while for $\fn \leq -1$ there are no poles at $z=0$. For $\fn=0$ there are poles both at $z=0$ and $z=\infty$, controlled by $A_n(z=0;\fm)$ and $A_n(z=\infty;\fm)$. However the series $\sum_{\fm\in\bZ}$ of such residues is convergent and can be resummed; moreover $\lim_{n\to\infty} A_n(z=0,\infty;\fm) = 0$. Therefore the contribution of the poles at $z=0,\infty$ to $I_n$ is smaller and smaller as $n$ is taken larger and larger, and can be neglected. In this case the integration contour can be deformed both towards zero or infinity.

Summarizing, we have shown that for
\be
|k| \leq \frac{|N_f - N_a|}2
\ee
$I_\infty$ can be computed by deforming the integration contour towards $z=0$ and/or $z=\infty$ and picking up the residues outside $z=0,\infty$, since the essential singularity at $z=0$ and/or $z=\infty$ does not contribute to the integral. For $|k|$ larger than the bound, the contributions from the essential singularities should be taken into account, although we will not try to do that here. Notice that exactly the same bound appeared in section \ref{sec: Rewriting Coulomb branch} when computing the $S^3$ partition function.

{
\bibliographystyle{utphys}
\bibliography{3dVortexLocalization}
}

\end{document}